\documentclass[onecolumn,floatfix,superscriptaddress,showpacs,showkeys,nofootinbib,preprint]{revtex4}     
\newcommand{\ave}[1]{\left\langle #1 \right\rangle}  
\usepackage{epsfig}   
\usepackage{amssymb,latexsym,amsmath}   

\begin{document}

\title{Multiplicity Fluctuations and Correlations in \\  
Limited Momentum Space Bins in Relativistic Gases}  
\author{Michael Hauer}   
\affiliation{Helmholtz Research School, University of Frankfurt, Frankfurt,   
  Germany}    
\affiliation{UCT-CERN Research Centre and Department of Physics,University of   
  Cape Town, Rondebosch 7701, South Africa}    
\author{Giorgio Torrieri}   
\affiliation{FIAS, J.W. Goethe Universit\"at,   
  Frankfurt am Main, Germany}   
\author{Spencer Wheaton}   
\affiliation{UCT-CERN Research Centre and Department of Physics,University of   
  Cape Town, Rondebosch 7701, South Africa}   
   
\begin{abstract}   
Multiplicity fluctuations and correlations are calculated within thermalized relativistic 
ideal quantum gases. These are shown to be sensitive to the choice of statistical ensemble 
as well as to the choice of acceptance window in momentum space. It is furthermore shown 
that global conservation laws introduce non-trivial correlations between disconnected 
regions in momentum space, even in the absence of any dynamics.
\end{abstract}   
   
\pacs{24.10.Pa, 24.60.Ky, 25.75.-q}  
  
\keywords{nucleus-nucleus collisions, statistical models,  
fluctuations}  
  
\maketitle   
  
\section{Introduction}   
\label{Intro}   
Fluctuations of, and correlations between, various experimental observables  
are believed to have the potential to reveal new physics.   
The growing interest in event-by-event fluctuations in strong interactions   
is motivated by expected anomalies in the vicinity of the onset of   
deconfinement \cite{OnsetOfDecon,koch1,asakawa,koch2} and in the case when the  
expanding system goes through the transition line between  
quark-gluon plasma and hadron gas \cite{PhaseTrans}. In  
particular, a critical point of strongly interacting matter may be  
accompanied by a characteristic power-law pattern in fluctuations  
\cite{CriticalPoint}. Recently, it has been suggested that correlations across a   
large interval of rapidity arise from a color glass condensate    
\cite{correlation_cgc1,correlation_cgc2}.  
In recent years a wide range of experimental measurements of fluctuations of particle   
multiplicities \cite{fluc-mult,BeniData}, transverse momenta \cite{fluc-pT} and   
multiplicity correlations in rapidity \cite{tarnowsky,PHENIX_rap_corr,PHOBOS_rap_corr}   
have been reported, leading to a lively discussion regarding their physical interpretation \cite{GMC,bzdak,correlation_cgc2,tarnowsky}.  
  
To get a reliable indication of new physics, it is important to note that most   
fluctuation and correlation observables are also sensitive to some ``baseline'' contributions   
that, nevertheless, can have non-trivial behaviour.  For instance, most fluctuation and   
correlation observables are sensitive to the global characteristics (e.g. the distribution
of the number of colliding nucleons)  of a sample of events, which can   
in turn be non-trivially constrained by centrality bin construction \cite{GMC}.  Similarly,    
conservation laws can provide a ``trivial'' correlation between observables.  The effects   
of such correlations depend on the {\em scale} at which these conservation   
laws become important. This scale could be anything, from microscopic (mean free path,   
diffusion scale) to the macroscopic size of the system.  
  
The purpose of this paper is to study these baseline correlations in a limiting case:   
that of a thermalized relativistic ideal (no inter-particle interactions) quantum gas for which we want to assess the importance   
of globally applied conservation laws for particle multiplicity fluctuations and correlations.   
In this case, all observables are calculable simply using statistical mechanics techniques. 
Such an approach has a long and distinguished history of calculating particle multiplicities in 
hadronic collisions ~\cite{Fer50,Pom51,Lan53,Hag65,bdm,equil_energy,jansbook,becattini,nuxu,share,sharev2,thermus}.
  
Conventionally in statistical mechanics three standard ensembles are   
discussed; the micro canonical ensemble (MCE), the canonical ensemble (CE),  
and the grand canonical ensemble (GCE). In the MCE\footnote{The term MCE is also often   
applied to ensembles with energy but not momentum conservation.} one considers an ensemble   
of micro states with exactly fixed values of extensive conserved quantities   
(energy, momentum, electric charge, etc.), with `a priori equal probabilities` of   
all micro states (see e.g. \cite{Patriha}). The CE introduces the concept of temperature by   
introduction of an infinite thermal bath, which can exchange energy (and momentum) with the system.  
The GCE introduces further chemical potentials by attaching the system under consideration   
to an infinite charge bath\footnote{Note that a system with many charges can have some   
charges described via the CE and others via the GCE.}.  
Only if the experimentally accessible system is only a small   
fraction of the total, and all parts have had the opportunity   
to mutually equilibrate, can the appropriate ensemble be the Grand Canonical one.  
  
In the limit of very large volume and constant density (the thermodynamic limit),   
average values of intensive quantities are the same for all ensembles.  However,   
even in this limit, these ensembles have different properties with respect to fluctuations   
and correlations \cite{SM_fluc_ce}. In the MCE, energy and charge are exactly fixed.   
In the CE, charge remains fixed, while energy is allowed to fluctuate about some average value.   
Finally in the GCE the requirement of exact charge conservation is dropped, too.  
One may also consider isobaric ensembles \cite{isobar}, or even more general `extended   
Gaussian ensembles` \cite{extgauss,alpha}. In previous articles  
\cite{SM_fluc_ce,CE_Res,MCEvsData,SM_fluc_mce,VdWFluc,QstatsNearTL,BoseCond,clt,acc,isobar,alpha}   
it was shown that these differences mean that multiplicity fluctuations are ultimately   
ensemble specific.   
  
In this article we extend these results to fluctuations and correlations between particle   
multiplicities  in limited bins of momentum space (rapidity $y$, transverse momentum   
$p_T$ and azimuthal angle $\phi$). In section \ref{Model} we present details   
of the calculation of correlations within statistical mechanics.  The following two   
sections present calculated fluctuations and correlations {\em within} the same momentum   
space bin for a stationary (section \ref{StaticSource}) and boosted   
(section \ref{BoostedSource}) system. In section \ref{LongRangeCorr} we discuss long   
range correlations {\em between} momentum space bins. A discussion section \ref{Summary}, summarising our results and discussing their phenomenological implications within the context of heavy ion collisions, closes the paper.  
  
\section{Correlations and Fluctuations within different Ensembles}   
\label{Model}   
  
In a recent paper \cite{clt} we have shown that GCE joint distributions of   
extensive quantities converge to Multivariate Normal Distributions (MND) in the thermodynamic   
limit (TL). MCE or CE multiplicity distributions could then be defined through   
conditional GCE distributions. In general one may write for the multiplicity distribution   
$P_{ce} (N)$ of a CE with conserved electric charge $Q$:  
\begin{equation}   
P_{ce} (N)~=~ \frac{\textrm{number of all states with $Q$ and   
    $N$}}{\textrm{number of all states with $Q$ }}~.  
\end{equation}     
Likewise one can write for the CE joint multiplicity distribution of particle species $A$   
and $B$:   
\begin{equation}   
P_{ce} (N_A,N_B)~=~ \frac{\textrm{number of all states with $Q$, $N_A$ and   
    $N_B$}}{\textrm{number of all states with $Q$ }}~.    
\end{equation}   
The number of all micro states with electric charge $Q$, and multiplicities $N_A$ and $N_B$  
of a system with temperature $T$ and volume $V$ is given by the CE partition function   
$Z_{CE}(V,T,Q,N_A,N_B)$. Similarly, $Z_{CE}(V,T,Q)$ denotes the number of micro states   
with fixed electric charge $Q$, but arbitrary multiplicities $N_A$ and $N_B$, for the   
same physical system.  
  
The strategy to calculate joint multiplicity distributions could thus be the following   
(in principle also valid at finite volume):  
\begin{eqnarray}   
\label{P_one}  
P_{ce} (N_A,N_B) &=& \frac{Z_{CE}(V,T,Q,N_A,N_B)}{Z_{CE}(V,T,Q)}~, \\   
\label{P_two}  
&=& \frac{e^{Q\frac{\mu}{T}}~Z_{CE}(V,T,Q,N_A,N_B)}{Z_{GCE}(V,T,\mu)} ~   
\frac{Z_{GCE}(V,T,\mu)}{e^{Q\frac{\mu}{T}}~Z_{CE}(V,T,Q)}~,  \\    
\label{P_three}  
&=& P_{gce}(Q,N_A,N_B) ~P_{gce}^{-1}(Q)  
~=~ P_{gce}(N_A,N_B|Q) ~.  
\end{eqnarray}   
In order to get from Eq.(\ref{P_one}) to Eq.(\ref{P_three}) both canonical partition functions   
$Z_{CE}(V,T,Q,N_A,N_B)$ and $Z_{CE}(V,T,Q)$ are divided by their GCE counterpart   
$Z_{GCE}(V,T,\mu)$ and multiplied by $e^{Q\frac{\mu}{T}}$.  
The first term on the right hand side of Eq.(\ref{P_two}) then equals the GCE joint   
distribution $P_{gce}(Q,N_A,N_B)$, while the second term is just the inverse of the GCE charge   
distribution $P_{gce}(Q)$. Their ratio is the (normalised) GCE conditional distribution  
 of particle multiplicities $N_A$ and $N_B$ at fixed electric charge $Q$, $P_{gce}(N_A,N_B|Q)$,   
and equals the CE distribution $P_{ce} (N_A,N_B)$ at the same value of $Q$. This result   
is independent of the choice of chemical potential $\mu$.  
  
The problem of finding a solution, or a (large volume) approximation, to the CE distribution   
$P_{ce} (N_A,N_B)$ is now turned into the problem of finding a solution or approximation to  
the GCE distribution of multiplicities $N_A$ and $N_B$, {\it and} charge $Q$.   
The role of chemical potential (or Lagrange multiplier) $\mu$ will be discussed in Section 
\ref{Results}.   
  
From the assumption that the GCE distribution $P_{gce}(Q,N_A,N_B)$ converges  
to a Tri-variate Normal Distribution, it also follows that the marginal distribution   
$P_{gce}(Q)$, as well as the conditional distribution $P_{gce}(N_A,N_B|Q) $, are   
Normal Distributions. Hence, $P_{ce} (N_A,N_B)$ should have a good approximation in a   
Bivariate Normal Distribution (BND) in the large volume limit (where the large particle numbers can be appropriately treated as continuous):  
\begin{eqnarray}\label{BND}   
P_{BND}(N_A,N_B) &=& \frac{1}{2\pi V \sqrt{\sigma_A^2 \sigma_B^2 \left(1-\rho^2   
    \right)}} \\   
&\times& \exp \Bigg[ -\frac{1}{2V} \Bigg[    
    \frac{\left(\Delta N_A\right)^2}{\sigma_A^2 \left(1-\rho^2 \right) }   
      - 2\rho \frac{\left(\Delta N_A \right)\left(\Delta N_B \right)}{\sigma_A   
        \sigma_B \left(1-\rho^2 \right) }    
        +  \frac{\left(\Delta N_B \right)^2}{\sigma_B^2 \left(1-\rho^2 \right)   
        } \Bigg] \Bigg]~, \nonumber   
\end{eqnarray}   
where $\Delta N_X = N_X - \langle N_X \rangle$, with $X=A,B$ and:   
\begin{eqnarray}   
V \sigma_A^2 &\equiv& \langle N_A^2 \rangle ~-~ \langle N_A \rangle^2~, \\   
V \sigma_B^2 &\equiv& \langle N_B^2 \rangle ~-~ \langle N_B \rangle^2~, \\    
V \sigma_{AB} &\equiv& \langle N_A N_B \rangle ~-~ \langle N_A \rangle \langle N_B   
\rangle~.    
\end{eqnarray}    
Here $V \sigma_A^2$ and $V \sigma_B^2$ are the variances of the marginal multiplicity   
distributions of particles $N_A$ and $N_B$. The term $V \sigma_{AB}$ is called the co-variance.  
Additionally we define the scaled variance $\omega_X$:  
\begin{equation}\label{omega}  
\omega_X ~=~ \frac{\langle N_X^2 \rangle ~-~ \langle N_X \rangle^2}{\langle N_X \rangle}~,  
\end{equation}  
which measures the width of the marginal distribution $P(N_X)$. Lastly,   
\begin{equation}\label{rho}  
\rho ~\equiv~ \frac{ \sigma_{AB}}{\sigma_A\sigma_B} ~,  
\end{equation}   
is the correlation coefficient between particle numbers $N_A$ and $N_B$.  
  
The distribution Eq.(\ref{BND}) hence has 5 parameters: Mean values   
$\langle N_A \rangle$ and $\langle N_B \rangle$, variances of marginal distributions   
$V \sigma_A^2$ and $V \sigma_B^2$, and the correlation coefficient $\rho$.   
Loosely speaking, the correlation coefficient $\rho$ defines how the BND, Eq.(\ref{BND}),   
is tilted. In the case where $\rho > 0$, the distribution is elongated along the main diagonal,   
and measuring a larger (smaller) number of particles $N_A$ implies that it is also more   
likely to measure a larger (smaller) number of particles $N_B$. The distribution is   
tilted the other way, if $\rho < 0$. In this case, multiplicities $N_A$ and $N_B$ are   
anti-correlated, and measuring $N_A > \langle N_A \rangle$ implies that it is now more   
likely to measure $N_B < \langle N_B \rangle$. Particle numbers $N_A$ and $N_B$ are   
uncorrelated, if $\rho = 0$.  
  
Similarly, we define MCE multiplicity distributions  
in terms of conditional GCE distributions of extensive quantities. For this we will   
first find a suitable approximation to the GCE joint distribution of extensive quantities   
(electric charge, energy, momentum, particle number(s), etc.) by Fourier analysis   
of the GCE partition function. The MCE multiplicity distribution  
is then given by a slice along a surface of constant values of extensive quantities.

\subsection{GCE Partition Function}   
\label{GCEPF}   
The GCE partition function of a relativistic gas with volume $V$,   
local temperature $T=1/\beta$, chemical potentials $\vec \mu$ and collective  
four velocity $\vec u$ reads (the system four-temperature \cite{fourtemp} is   
$\vec \beta~=~\beta \vec u$):    
\begin{equation}\label{ZGCE}  
Z_{GCE}(V,\beta,\vec \mu,\vec u)~=~ \exp \Bigg[V \Psi   
\left(\beta,\vec \mu,\vec u \right) \Bigg]~,   
\end{equation}   
where $\Psi \left( \beta,\vec \mu,\vec u \right)$ is a sum over the single particle  
partition functions $\psi_l \left(\beta,\vec \mu,\vec u \right)$ of all particle   
species `$l$` considered in the model:  
\begin{equation} \label{Psi}  
\Psi \left( \beta,\vec \mu,\vec u \right) ~=~ \sum_{l} ~\psi_l   
\left( \beta,\vec \mu,\vec u \right)~.  
\end{equation}   
The single particle partition function $\psi_l \left(\beta,\vec \mu,\vec u \right)$   
of particle species `$l$` is given by a J\"uttner distribution:  
\begin{equation} \label{psi}  
\psi_l \left(\beta,\vec \mu,\vec u \right) ~=~   
\frac{g_l}{\left(2 \pi \right)^3} ~\int d^3p ~  
\ln \left(1~\pm~e^{-\beta ~p_l^{\mu}~u_{\mu} ~+~ \beta ~q_l^j~ \mu_j}   
\right)^{\pm1}~,  
\end{equation}   
where $p_l^{\mu}$ are the components of the four momentum, $q_l^j$ are the components of   
the charge vector and $g_l$ is the degeneracy factor. The upper sign refers to Fermi-Dirac   
statistics (FD), while the lower sign refers to Bose-Einstein statistics (BE). The case   
of Maxwell-Boltzmann (MB) statistics is analogous.  
  
In the following we restrict ourselves to systems moving along the $z$-axis   
and use variables $y$, $p_T$ and~$\phi$. For a boost in rapidity of $y_0$ one finds   
for the four-velocity $\vec {u} $, the four-momentum $\vec p_l $ and the integral measure   
$d^3p$, respectively:   
\begin{eqnarray}   
\label{vec_u}  
\vec {u} &=& \left( \cosh y_0,~ 0,~0,~\sinh y_0 \right)~,\\   
\vec p_l &=& \left( \sqrt{m_l^2+p_T^2} \cosh y,~ p_T \cos \phi ,  
~ p_T \sin \phi, ~\sqrt{m_l^2+p_T^2} \sinh y \right)~,\\   
d^3p &=& p_T~ \sqrt{m_l^2+p_T^2} ~ \cosh (y-y_0)~dy~dp_T~d\phi~,  
\end{eqnarray}   
where $m_l$ is the mass of a particle of species $l$.  
The single particle partition function Eq.(\ref{psi}) now reads:  
\begin{equation}   
\psi_l \left(\beta,\vec \mu,\vec u \right) ~=~   
\frac{g_l}{\left(2 \pi \right)^3}   
\int \limits_{-\pi}^{\pi} d\phi \int \limits_0^{\infty} dp_T    
\int \limits_{-\infty}^{\infty} dy ~  
p_T \sqrt{m_l^2+p_T^2} \cosh (y-y_0)   
~\ln \left(1~\pm~e^{-\beta ~p_l^{\mu}~u_{\mu} ~+~ \beta ~q_l^j~ \mu_j} \right)^{\pm1}.  
\end{equation}    
For the examples in the following sections we chose a simple gas with only one conserved charge,   
denoted as a `pion gas`. The presented formulae are, however, also readily applicable to a   
hadron resonance gas (HRG).  
Depending on what system one may want to study, one introduces chemical   
potentials $\vec \mu $ and the `charge` vector $\vec q_l$ of particle species $l$:  
\begin{eqnarray}  
\vec \mu &=& (\mu_B,~\mu_S,~\mu_Q,~ \mu_{N_A},~ \mu_{N_B}) \qquad   
\vec q_l = (b_l,~s_l,~q_l,~ n_A(\Omega),~ n_B(\Omega)) \qquad  
\textrm{for a HRG} ~,\\    
\vec \mu &=& (\mu_{Q},~ \mu_{N_A},~ \mu_{N_B}) \qquad \qquad \qquad   
\vec q_l = (q_l,~ n_A(\Omega),~ n_B(\Omega)) \qquad \qquad    
\textrm{for a pion gas}~,  
\end{eqnarray}  
where $\mu_B$, $\mu_S$, and $\mu_Q$ are the baryon, strangeness,   
and electric charge chemical potentials, respectively. $\mu_{N_A}$ and $\mu_{N_B}$   
are particle-specific chemical potentials, and could denote out of chemical   
equilibrium multiplicities of species `A` and `B`, similar  
to phase space occupancy factors $\gamma_S$ \cite{gammaSfirst} and $\gamma_q$   
\cite{gammaQfirst}. Throughout this paper we neglect finite density effects,   
so $ \mu_{N_A} =  \mu_{N_B} = 0$.  
  
In addition, $b_l$, $s_l$, and $q_l$ are the baryonic charge, the strangeness, and the   
electric charge of particle species `$l$`. $\Omega$ is the momentum space bin in which we   
wish to measure particle multiplicity. $n_A(\Omega) = 1$ if the momentum vector of the   
particle is within the acceptance, $n_A(\Omega) = 0$ if not. The charge vector $\vec q_l$   
also contains, to maintain a common notation for all particle species considered in  
Eq.(\ref{Psi}), the `quantum` number $n_B(\Omega)$.   
  
One may also be interested in correlations of, for instance,  
 baryon number $B$ and strangeness $S$, as e.g. in Refs.\cite{QCD_Karsch,Koch_hadronic_fluc}.   
In this case, the $\Lambda$ particle, with $q_{\Lambda}=(1,-1,0,1,1)$, would be counted in   
groups $A$ and $B$, provided the momentum vector is within the acceptance $\Omega$.

\subsection{Generating Function}   
\label{MCEPF}   
To introduce the generating function of the charge distribution   
$Z_{GCE}(V,\beta,\vec \mu,\vec u; \vec \phi, \vec \alpha)$   
in the GCE we substitute in Eq.(\ref{psi}):  
\begin{eqnarray} \label{subst_1}  
\beta ~\mu_j &\rightarrow& \beta ~\mu_j ~+~ i \phi_{j}~, \\ \label{subst_2}  
\beta ~u_{\mu} &\rightarrow& \beta ~u_{\mu} ~-~ i \alpha_{\mu} ~.  
\end{eqnarray}   
The yet unnormalised joint probability distribution of extensive   
quantities $\vec Q,\vec P$ in the GCE is then given by the Fourier   
transform of Eq.(\ref{ZGCE}) after substitutions Eqs.(\ref{subst_1},\ref{subst_2}):  
\begin{equation} \label{Curly}  
\mathcal{Z}^{\vec Q,\vec P}(V,\beta,\vec \mu,\vec u) ~=~   
\int \limits_{-\pi}^{\pi} \frac{d^J\phi}{\left( 2\pi \right)^J} ~e^{-iQ^j \phi_{j}} ~   
\int \limits_{-\infty}^{\infty} \frac{d^4\alpha }{\left( 2\pi \right)^4}    
~ e^{-iP^{\mu} \alpha_{\mu}}   
~\exp \Bigg[V \Psi \left(\beta,\vec \mu,\vec u ; \vec \phi, \vec \alpha \right) \Bigg]~.   
\end{equation}   
More details of the calculation, in particular on the connection between the partition  
functions $\mathcal{Z}^{\vec Q,\vec P}(V,\beta,\vec \mu,\vec u)$ and the conventional   
version $Z_{MCE}(V,\vec Q,\vec P)$ \cite{fourtemp,fourtempII}, can be found in   
Appendix \ref{MCEPF}. Depending on the system under consideration, we introduce the vector of  
extensive quantities $\vec Q$ and corresponding Wick rotated fugacities $\vec \phi$:   
\begin{eqnarray}   
\vec Q &=& (B,~S,~Q, N_A, N_B) \qquad   
\vec \phi = (\phi_B,~\phi_S,~\phi_Q,~ \phi_{N_A},~ \phi_{N_B}) \qquad   
\textrm{for a HRG}~, \\   
\vec Q &=& (Q, N_A, N_B) \qquad \qquad \quad  
\vec \phi = (\phi_{Q},~ \phi_{N_A},~ \phi_{N_B}) \qquad \qquad \quad  
\textrm{for a pion gas}~.  
\end{eqnarray}   
Here $B$ is the baryon number, $S$ is the strangeness, and $Q$ is the electric charge   
of the system. Together with particle numbers $N_A$ and $N_B$ this would be a 5-dimensional  
distribution in the case of a CE HRG. Additionally for four-momentum conservation,   
yielding a 9-dimensional Fourier transform Eq.(\ref{Curly}) for a MCE HRG, we write:  
\begin{equation}  
\vec P ~=~ (E,~P_x,~P_y,~P_z) \qquad  
\vec \alpha ~=~ (\alpha_{E},~\alpha_{P_x},~\alpha_{P_y},~\alpha_{P_z})~,  
\end{equation}   
where $E$ is the energy and $P_{x}$, $P_{y}$, and $P_{z}$ are the components of the   
collective momentum of the system, while $\vec \alpha$ are the corresponding fugacities.  
  
The integrand of Eq.(\ref{Curly}) is sharply peaked at the origin   
$\vec \phi = \vec \alpha = \vec 0$ in the TL \cite{clt}. The main contribution therefore  
comes from a very small region. To see this, a second derivative test can be done on   
the integrand of Eq.(\ref{Curly}) taking into account the first two terms of Eq.(\ref{kappa}).   
The limits of integration can hence be extended to $\pm \infty$. The distinction between   
discrete (Kronecker $\delta$) and continuous quantities (Dirac $\delta$) is not relevant for   
the TL approximation, where particle number is a continuous variable to be integrated over. We thus proceed by Taylor expansion of Eq.(\ref{Psi}). For this it   
is convenient to include everything into a common vector notation:  
\begin{equation}  
\vec{\mathcal{Q}} ~=~ \left( \vec Q,~ \vec P \right) \qquad \textrm{and} \qquad  
\vec \theta ~=~ \left( \vec \phi,~ \vec \alpha \right)~.  
\end{equation}  
The dimensionality of the vector $\vec{\mathcal{Q}}$ is denoted as $J=2+3+4=9$ for a MCE HRG.  
We now expand the cumulant generating function,   
$\Psi \left(\beta,\vec \mu,\vec u ; \vec \theta \right)$,   
in a Taylor series:  
\begin{equation}\label{Psi_expand}  
\Psi \left(\beta,\vec \mu,\vec u ; \vec \theta \right)   
~\simeq~ \sum \limits_{n=0}^{\infty} ~ \frac{i^n}{n!}~  
\kappa_n^{j_1,j_2, \dots, j_n}~   
\theta_{j_1} ~ \theta_{j_2} \dots  \theta_{j_n}~,  
\end{equation}  
where the elements of the cumulant tensor, $\kappa_n^{j_1,j_2,\dots, j_n}$, are defined by:  
\begin{equation}\label{kappa}  
\kappa_n^{j_1,j_2,\dots, j_n}~ = \left( -i \right)^n~  
\frac{\partial^n \Psi}{\partial \theta_{j_1} \partial \theta_{j_2} \dots \partial \theta_{j_n}}   
\Bigg|_{\vec \theta = \vec 0}~.  
\end{equation}  
Generally cumulants are tensors of dimension $J$ and order $n$. The first cumulant is then   
a vector, while the second cumulant is a symmetric $J \times J$ matrix.  
A good approximation to Eq.(\ref{Curly}) around the point   
$\vec{\mathcal{Q}}_{eq} = (\vec Q_{eq}$, $\vec P_{eq})$,   
can be found in terms of a Taylor expansion of Eq.(\ref{Psi}) in $\vec \theta = ( \vec \phi, \vec   
\alpha )$, if:  
\begin{equation} \label{LM_mu}   
\frac{\partial \mathcal{Z}^{\vec {\mathcal{Q}}}(V,\beta,\vec \mu,\vec u)}{\partial   
  \vec{\mathcal{Q}}} \Bigg|_{\vec{\mathcal{Q}}_{eq}}~=~ 0 ~.  
\end{equation}   
  
Implicitly, Eq.(\ref{LM_mu}) does not define chemical potentials $\vec \mu$   
and four-temperature $\vec \beta = \beta \vec u$, but corresponding Lagrange multipliers   
which maximise the amplitude of the Fourier spectrum of the generating function   
for a desired value of $\vec Q_{eq}$ and $\vec P_{eq}$. Their values generally differ from   
the GCE set $(\beta,\vec \mu,\vec u)$, however they  coincide in the TL. Lagrange multipliers   
can be used for finite volume corrections \cite{clt}. In the following we   
restrict ourselves to the large volume approximation.  
  
\subsection{Joint Distributions}   
\label{GCEJD }   
  
In the large volume limit, i.e. $V \rightarrow \infty$, one may use the asymptotic solution,  
and only consider the first two cumulants, Eq.(\ref{kappa}). Substituting   
Eq.(\ref{Psi_expand}) into Eq.(\ref{Curly}) yields a standard $J$ dimensional Gaussian   
integral with solution:  
\begin{equation}\label{MND}   
 \mathcal{Z}^{\vec {\mathcal{Q}}}(V,T,\vec \mu, \vec u) ~\simeq~ Z_{GCE}(V,\beta,  
\vec \mu, \vec u)   
~ \frac{1}{\left(2\pi V\right)^{J/2}} ~\frac{1}{\det   
  \sigma} ~\exp \left[-\frac{1}{2}~ \xi^j \xi_j \right]~,  
\end{equation}   
where the elements of the new variable $\vec \xi$ are defined by:  
\begin{equation}\label{xi}   
\xi^j ~=~ \left( \Delta \mathcal{Q} \right)^k ~ \left( \sigma^{-1} \right)_k^j ~V^{-1/2}~.  
\end{equation}   
The elements of the vector $( \Delta \vec{\mathcal{Q}} )$ measure the distance of  
the charge vector $\vec{\mathcal{Q}}$ to the GCE mean $V \vec \kappa_1$:   
\begin{equation}\label{DQ_k}   
\left( \Delta \mathcal{Q} \right)^k ~\equiv~ \left(\mathcal{Q}^k - V \kappa_1^k \right)~,  
\end{equation}   
and $\left( \sigma^{-1} \right)$ is the inverse square root of the second   
order cumulant $\kappa_2$:  
\begin{equation}\label{invsigma}   
\sigma^{-1} ~\equiv~ \kappa_2^{-1/2}~.  
\end{equation}  
The GCE joint distribution of extensive quantities $\vec{\mathcal{Q}}$ is a   
MND\footnote{Finite   
volume corrections to Eq.(\ref{GCE_dist}) converge like $V^{-1/2}$ in the TL \cite{clt}.}:  
\begin{equation}\label{GCE_dist}  
P_{gce} (\vec{\mathcal{Q}}) ~=~ \frac{ \mathcal{Z}^{\vec {\mathcal{Q}}}(V,\beta,\vec \mu, \vec u)}{Z_{GCE}(V,\beta,\vec \mu, \vec u) } ~\simeq~   
~ \frac{1}{\left(2\pi V\right)^{J/2}} ~\frac{1}{\det   
  \sigma} ~\exp \left[-\frac{1}{2}~ \xi^j \xi_j \right]~.  
\end{equation}  
Mean values in the TL are given by the first Taylor expansion terms, $\langle N_A \rangle  
= V \kappa_1^{N_A}$, $\langle Q \rangle = V \kappa_1^{Q}$, $\langle E \rangle  
= V \kappa_1^{E}$, etc. and converge to GCE values. To obtain a joint (two-dimensional) particle   
multiplicity distribution one has to take a two-dimensional slice of the ($J$ dimensional)   
GCE distribution, Eq.(\ref{GCE_dist}), around the peak of the   
extensive quantities which one is considering fixed.   
The co-variance tensor $\kappa_2$ will be spelled out explicitly and discussed in   
Section \ref{BoostedSource}. Further details of the calculation can be found in Appendix   
\ref{AJD}.  
   
\section{Fluctuations and Correlations within a Momentum Bin}   
\label{Results}   
\subsection{Static System}   
\label{StaticSource}   
Let us start by discussing properties of a static thermal system. We want to measure joint   
distributions of multiplicities $N_A$ and $N_B$ in limited bins of momentum space   
(of width $\Delta p_T$ for transverse momentum bins and $\Delta y$ for rapidity bins).  
Depending on the size and positions of the bins, one finds different fluctuations and   
correlations.   
Results will, in particular, be compared to the acceptance scaling approximation employed in   
\cite{MCEvsData,CE_Res,SM_fluc_ce},   
which assumes random observation of particles with a certain probability $q$, regardless   
of particle momentum (see also Appendix \ref{AccS}). Corresponding results   
for scaled variance in MB statistics can also be found in~\cite{acc}.   
  
\begin{figure}[ht!]   
\epsfig{file=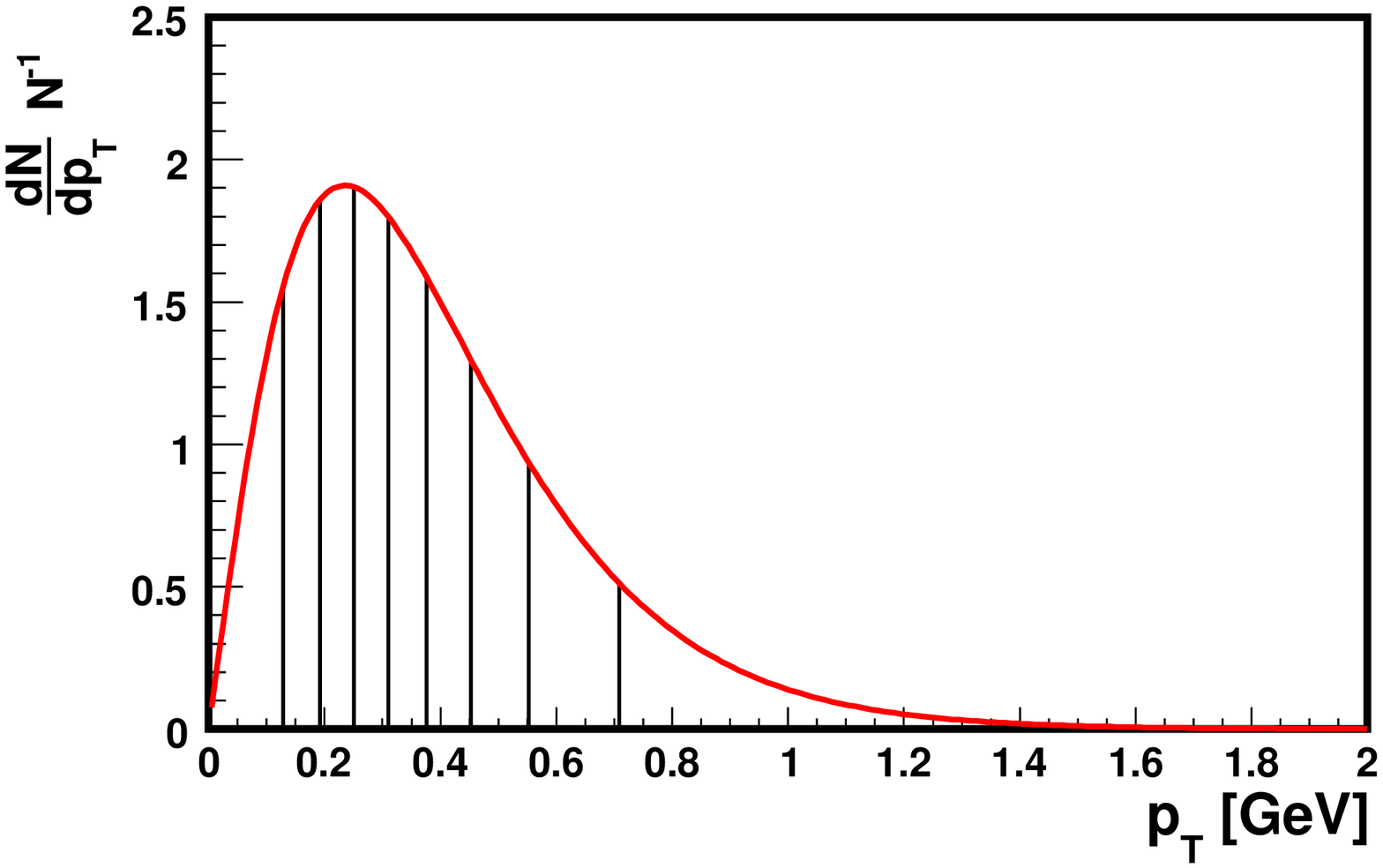,width=8.9cm}   
\epsfig{file=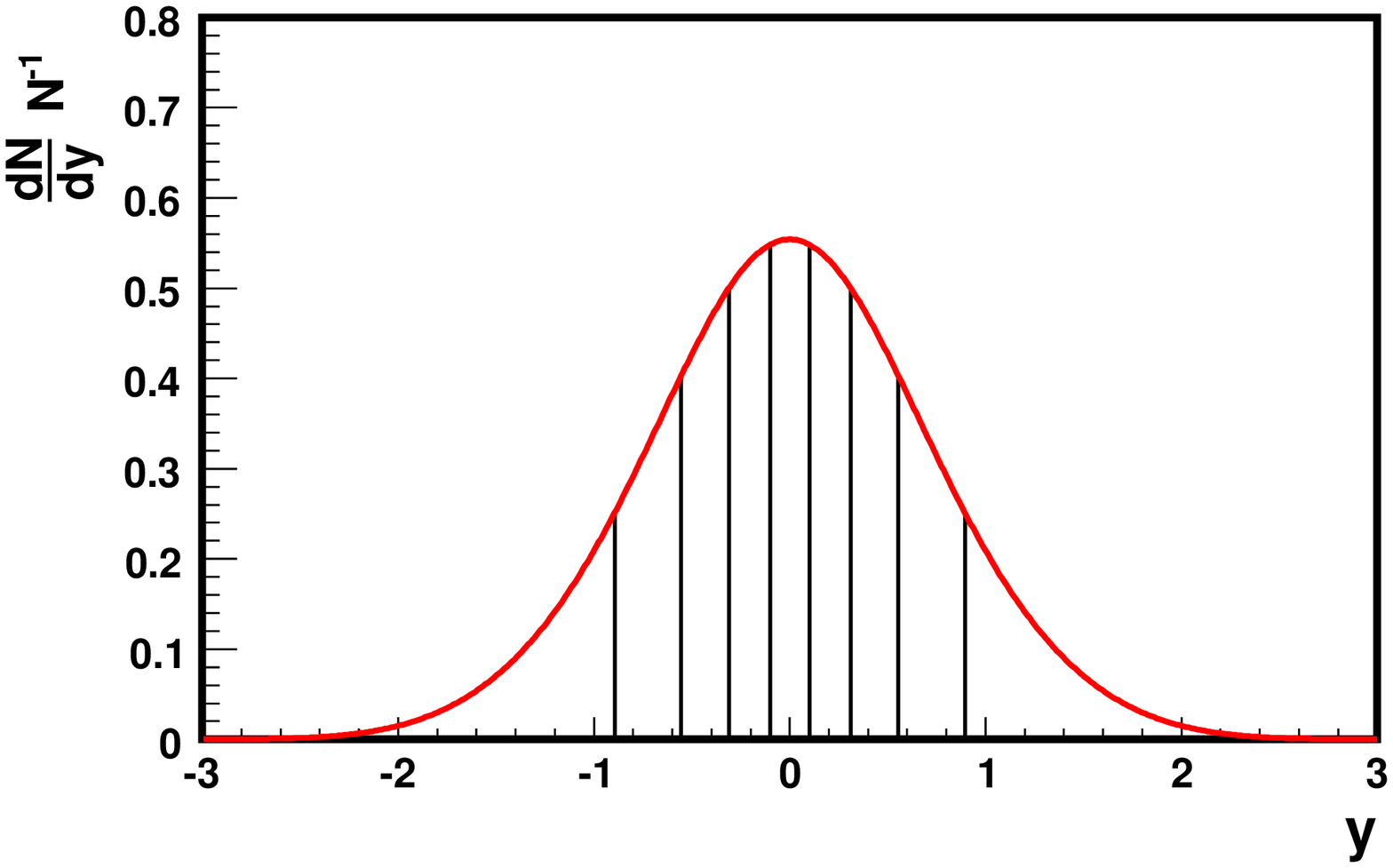,width=8.9cm}   
\caption{(Color online) Differential particle spectra for a `pion gas' at   
  $T=160$MeV. Transverse momentum spectrum ({\it left}) and    
  rapidity spectrum ({\it right}). Both curves are normalised to unity.   
  The bins are constructed such that each bin  contains  $1/9$ of the total yield.}    
\label{Boltzmannspectra}   
\end{figure}

For our examples we chose a gas with three degenerate massive particles   
(with positive, negative and zero charge) with mass ${m=0.140}$GeV in three   
different statistics (MB, FD, BE). The momentum spectra are assumed to be ideal GCE spectra,   
due to the large volume approximation. In Fig.(\ref{Boltzmannspectra}) transverse   
momentum and rapidity spectra are shown for MB statistics. BE and FD statistics yield   
similar spectra, unless chemical potentials are large.   
  
We can then define bins by requiring each bin to hold the same fraction of the total   
multiplicity. Note that in this case the width and position of bins $\Delta p_T$ and   
$\Delta y$ will strongly depend on the underlying momentum spectra. Our examples, in 
particular the FD case, are a little academic in the sense that there is no fermion of this
mass. In a HRG, often applied to heavy ion collisions, the lightest fermion is the nucleon
for which quantum effects are probably negligible.
   
\begin{figure}[ht!]   
\epsfig{file=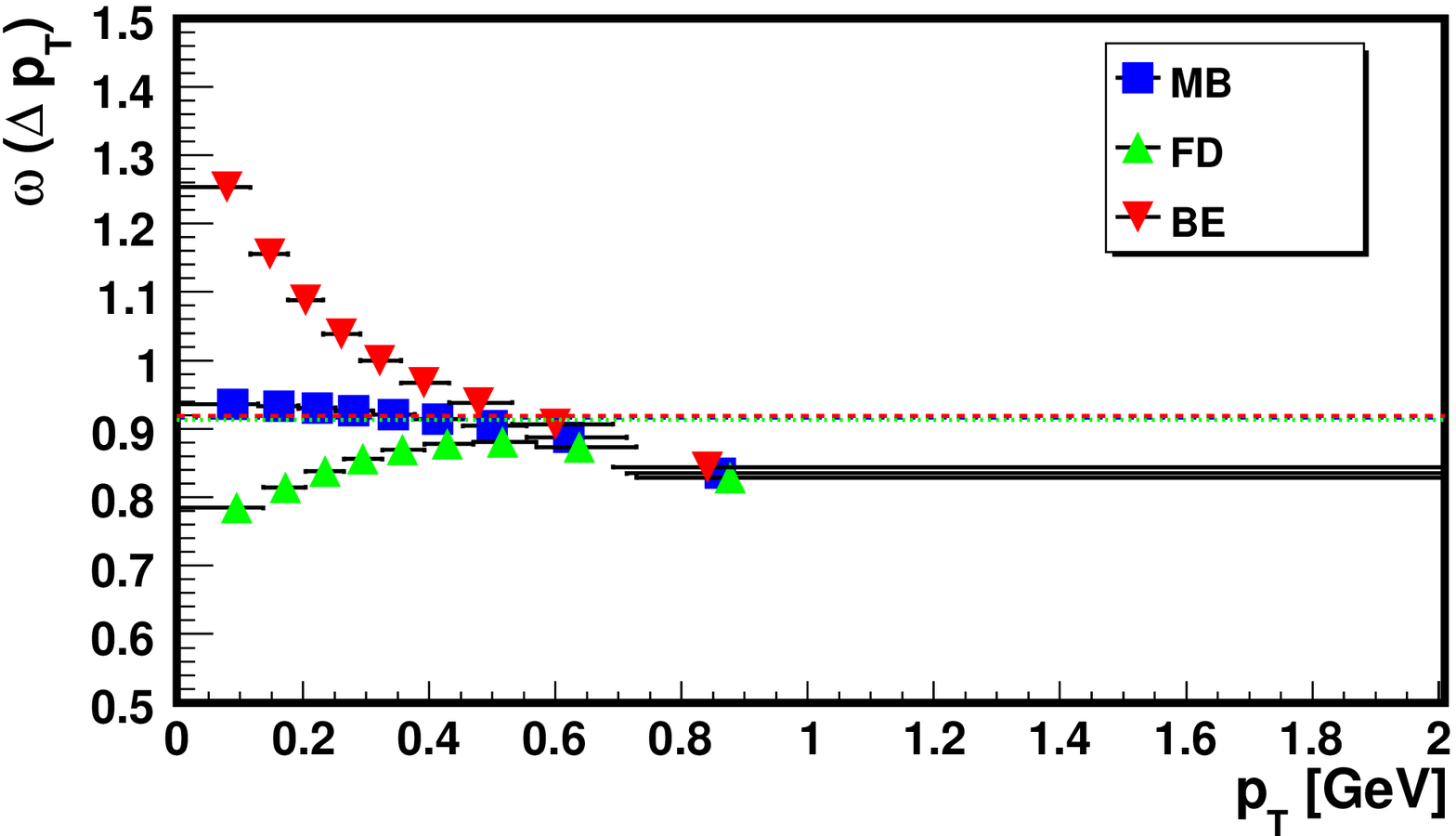,width=8.9cm,height=5.5cm}   
\epsfig{file=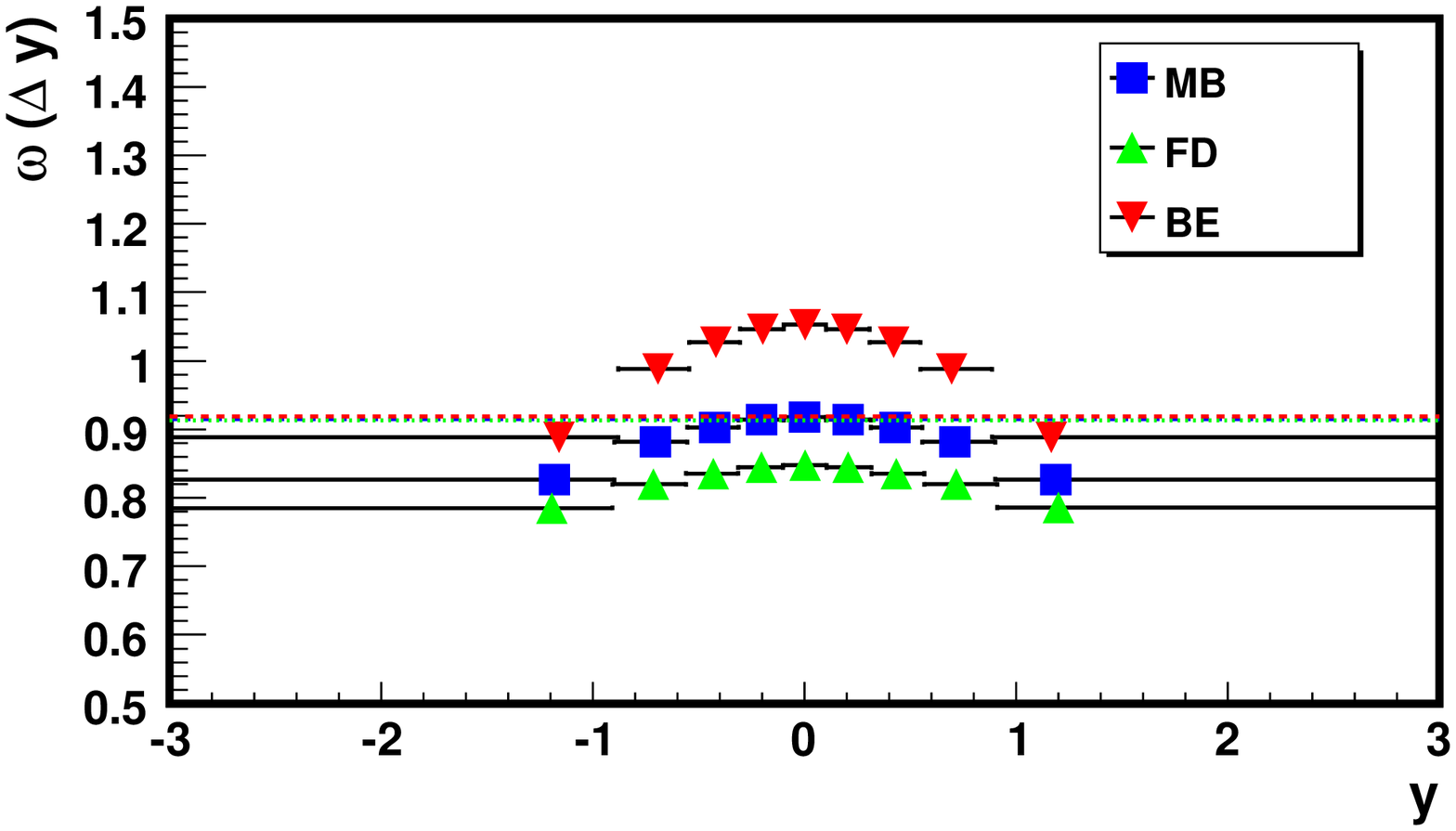,width=8.9cm,height=5.5cm}   
\caption{(Color online) Transverse momentum ({\it left}) and rapidity dependence ({\it right})  
  of the MCE scaled variance of negatively charged particles at {$T=160$MeV}, for a MB (blue),   
  FD (green), BE (red) `pion gas'  at zero charge density. Momentum bins are constructed   
  such that each bin contains the same fraction $q$ of the average $\pi^-$   
  yield. The horizontal bars indicate the width of the $\Delta p_T$ or $\Delta y$ bins,  while   
  the marker indicates the position of the center of gravity of the   
  corresponding bin. Dashed lines indicate acceptance scaling results, Eq.(\ref{acc_omega}).}     
\label{StaticSource_dodp}   
\end{figure}

In Fig.(\ref{StaticSource_dodp}) we present the scaled variance $\omega$, calculated   
using Eq.(\ref{omega}), within different transverse momentum bins   
$\Delta p_T$ ({\it left}) and rapidity bins $\Delta y$ ({\it right}).  
The scaled variance in limited bins of momentum space is more sensitive   
to the choice of particle statistics than the spectra would suggest. BE and FD effects   
 are particularly strong in momentum space bins in which occupation numbers are large. 
Hence, at the low momentum tail one finds suppression of fluctuations for FD and enhancement 
for BE, while at the high momentum tail, one finds 
$\omega_{BE} \simeq \omega_{MB} \simeq \omega_{FD}$, Fig.(\ref{StaticSource_dodp}) ({\it left}). 
The rapidity dependence,   
Fig.(\ref{StaticSource_dodp}) ({\it right}), has a different behaviour. The reason is, 
that in any $\Delta y$ bin there is some contribution from a low $p_T$ tail of the 
differential momentum spectrum $dN / dy / dp_{T}$ where quantum statistics effects 
are pronounced. This leads to a clear separation of the curves and one finds 
$\omega_{BE}~>~\omega_{MB}~>~\omega_{FD}$.
In contrast to this, the $4 \pi$ integrated (all particles observed) scaled variance   
is rather insensitive to the choice of statistics \cite{QstatsNearTL}  
(unless chemical potentials are large). Please note that there are in fact 3 different 
`acceptance scaling` lines in Fig.(\ref{StaticSource_dodp}), which extrapolate the $4 \pi$ 
integrated scaled variance to limited acceptance. The differences are however very small 
and all 3 lines lie practically on top of each other.       
  
\begin{figure}[ht!]   
\epsfig{file=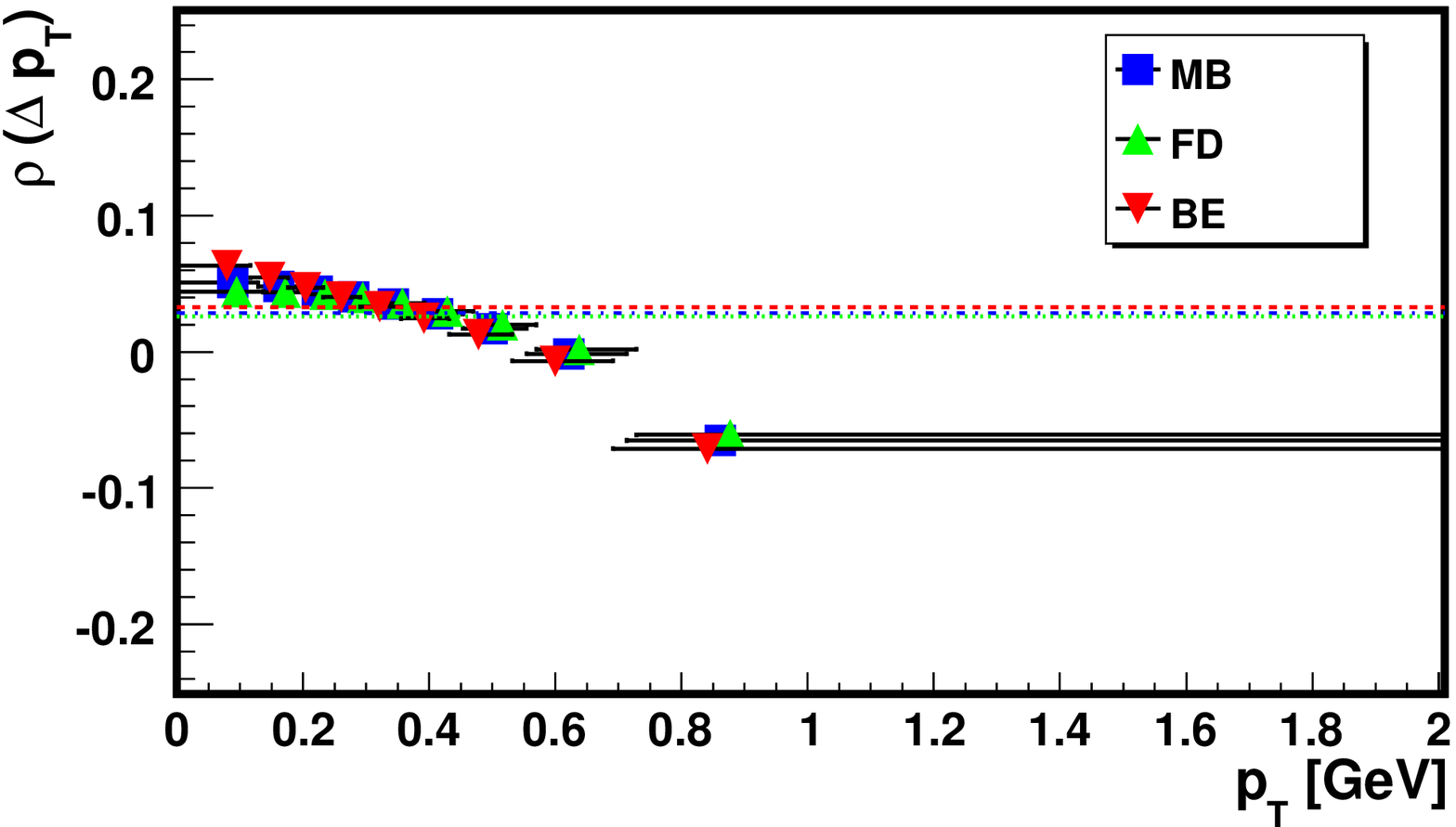,width=8.9cm,height=5.5cm}   
\epsfig{file=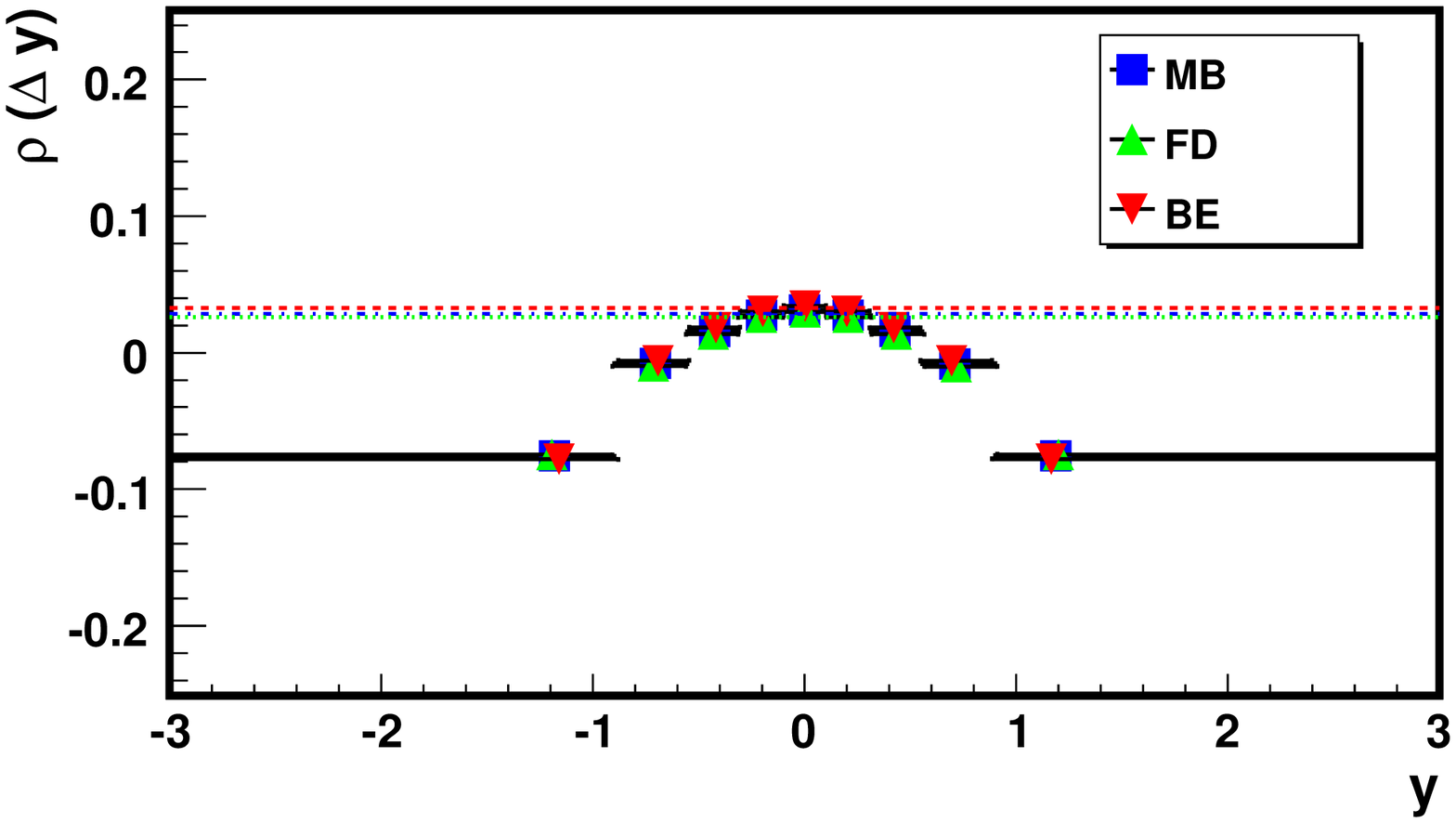,width=8.9cm,height=5.5cm}   
\caption{(Color online) Same as Fig.(\ref{StaticSource_dodp}), but for the MCE correlation   
  coefficient between positively and negatively charged particles.  
  Dashed lines indicate acceptance scaling results, Eq.(\ref{acc_rho}).}     
\label{StaticSource_drhodp}   
\end{figure}

In Fig.(\ref{StaticSource_drhodp}) we present the correlation coefficient $\rho$, calculated   
using Eq.(\ref{rho}), between positively and negatively charged particles in transverse   
momentum bins $\Delta p_T$ ({\it left}) and rapidity bins $\Delta y$ ({\it right}).   
The $4 \pi$ integrated correlation coefficient between positively and negatively charged   
particles would be $\rho_{4\pi} =1$ in the CE and MCE. In the GCE it would be $0$.   
In the MB CE it would not show any momentum space dependence and would   
always be $\rho>0$. In the MCE the situation is qualitatively different: in low   
momentum bins particles are positively correlated, while in high momentum bins they   
can even be anti-correlated. Horizontal lines again indicate acceptance scaling    
(Appendix~\ref{AccS}). Quantum effects for the correlation coefficient remain small as there 
is no explicit local (quantum) correlations between particles of different charge.  
   
\begin{figure}[ht!]   
\epsfig{file=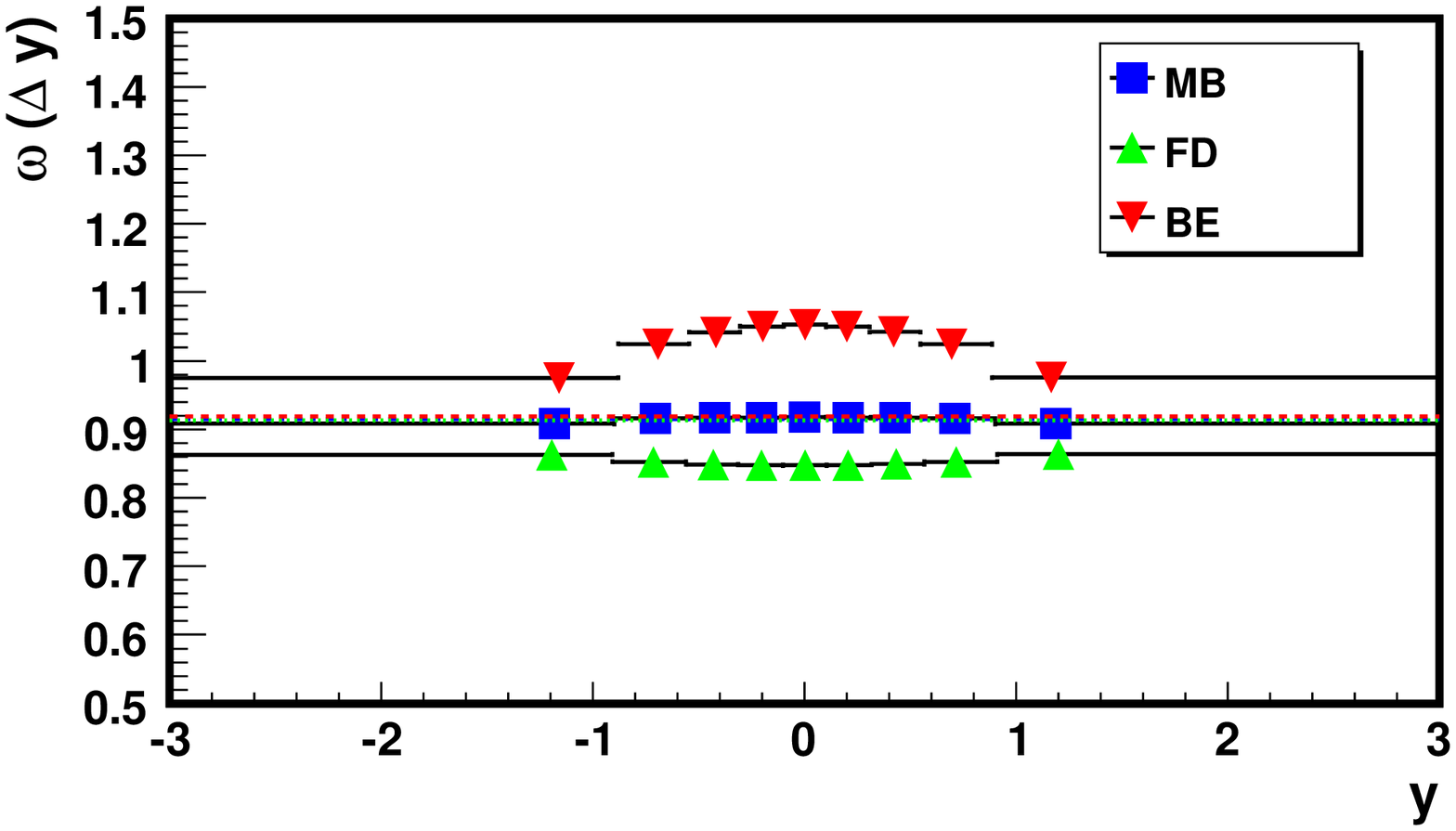,width=8.9cm,height=5.5cm}   
\epsfig{file=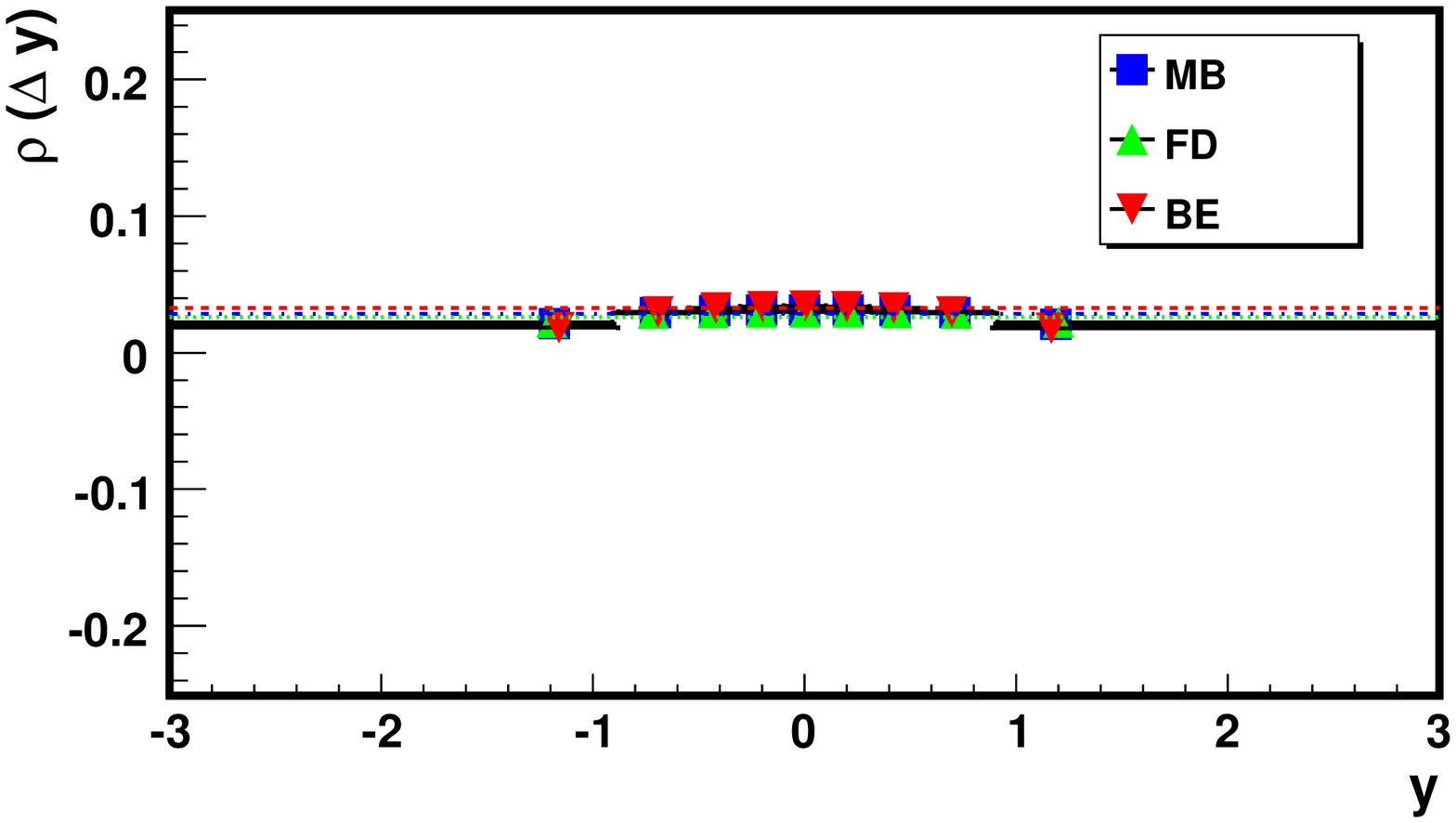,width=8.9cm,height=5.5cm}   
\caption{(Color online) Rapidity dependence of the scaled variance of positively charged particles ({\it left})
and of the correlation coefficient between positively and negatively charged particles ({\it right}).
Calculations are done for the same system as in 
Figs.(\ref{StaticSource_dodp},\ref{StaticSource_drhodp}), 
however disregarding momentum conservation.}     
\label{StaticSource_dxdy_nopz}   
\end{figure}

It should be stressed that the $\Delta p_T$ dependence in   
Figs.(\ref{StaticSource_dodp},\ref{StaticSource_drhodp}) is a direct consequence  
of energy conservation. The $\Delta y$ dependence of $\omega$ and $\rho$, however, is due  
to joint energy and longitudinal momentum ($P_z$) conservation. Disregarding $P_z$   
conservation leads to a substantially milder $\Delta y$ dependence, 
see Fig.(\ref{StaticSource_dxdy_nopz}).  

This behaviour can be intuitively explained: in a low momentum bin it is comparatively easy  
to balance charge, as each individual particle carries little energy and momentum. In contrast to   
this, in a high momentum bin with, say an excess of positively charged particles, it is   
unfavourable to balance charge, as one would also have to have more than on average   
negatively charged particles, and each particle carries large energy and momentum.   
This leads to suppressed fluctuations and correlations in high momentum bins when   
compared to low momentum bins.   
   
\begin{figure}   
\epsfig{file=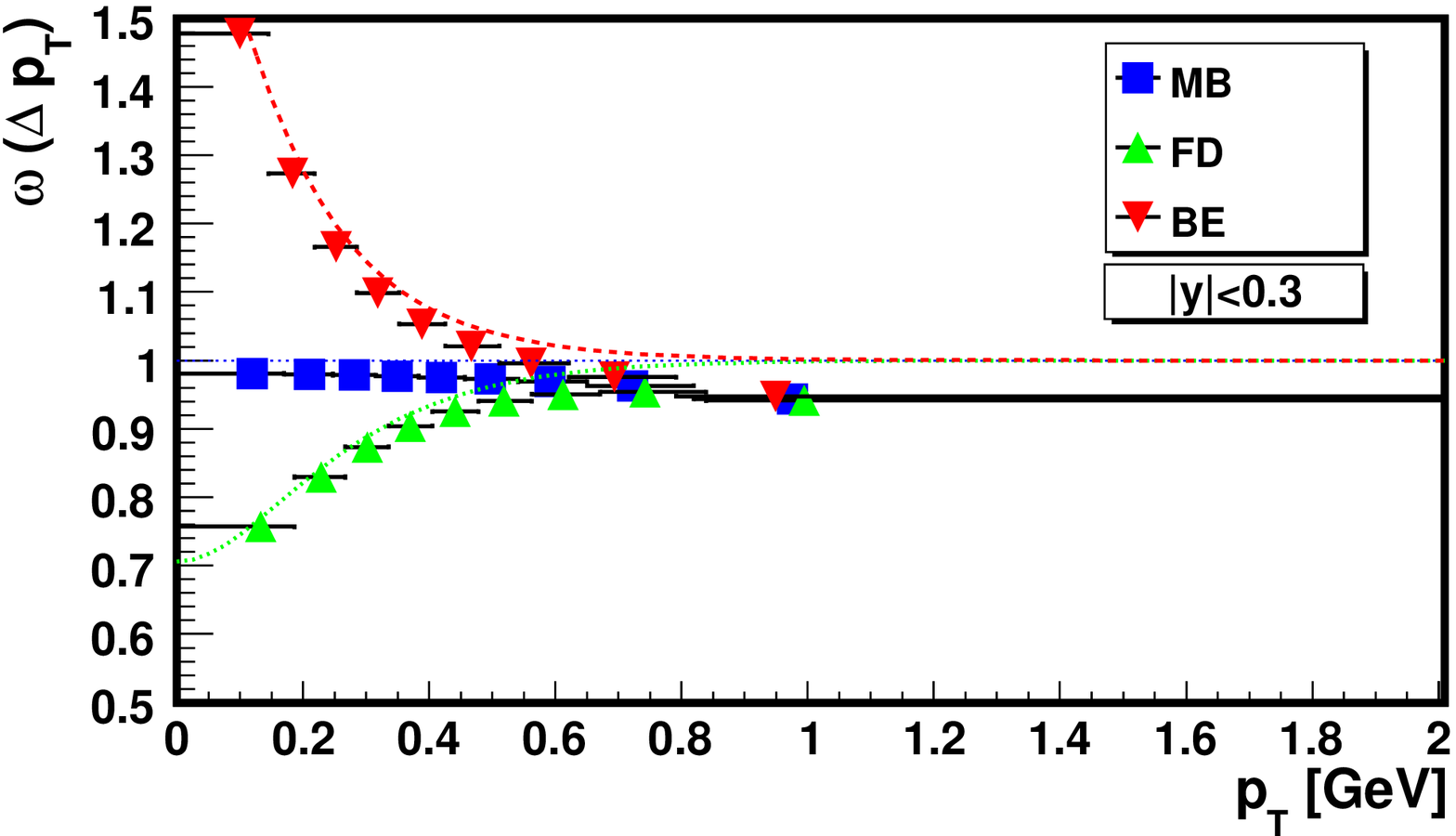,width=8.9cm,height=5.5cm}   
\epsfig{file=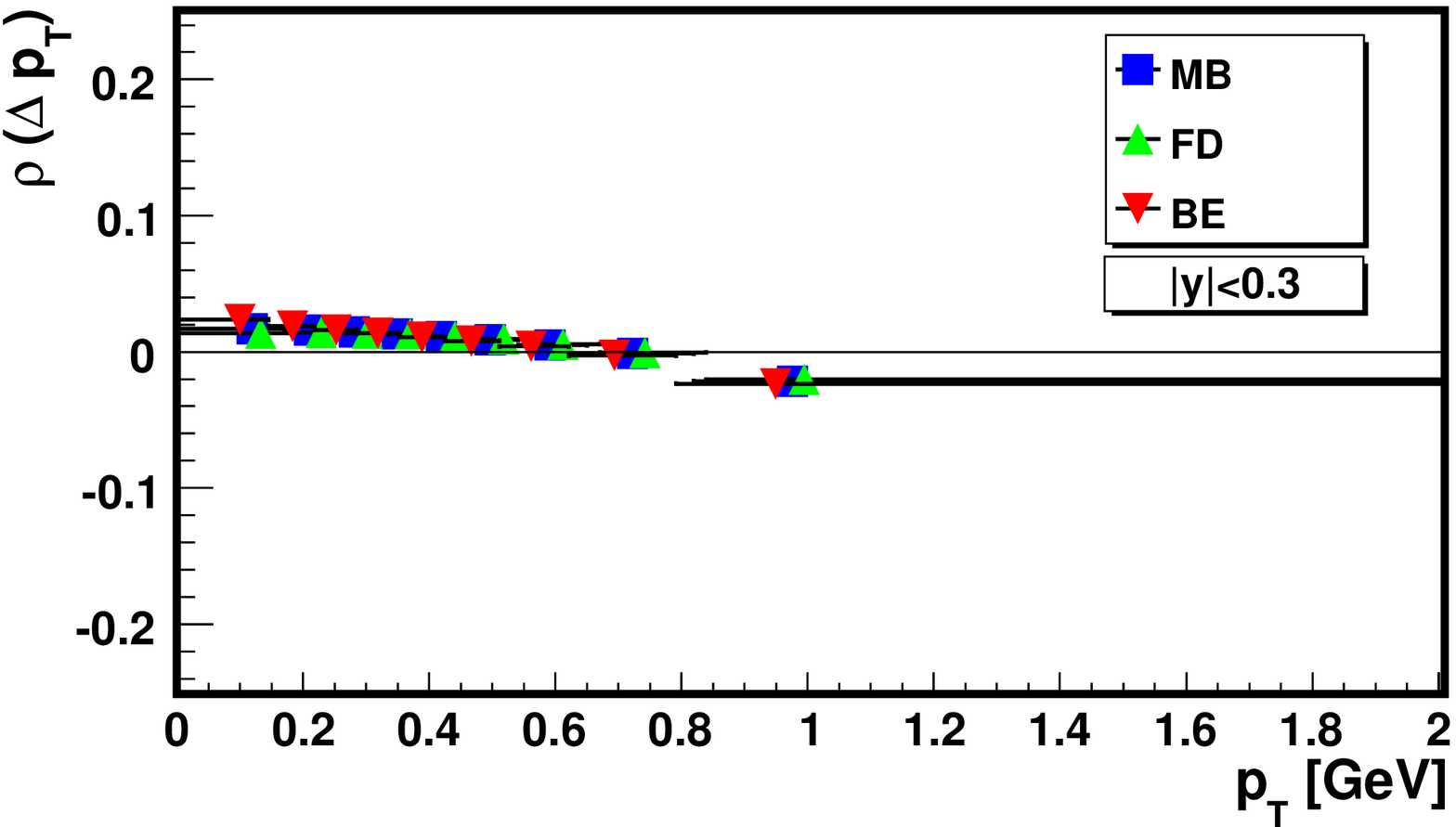,width=8.9cm,height=5.5cm}   
\caption{(Color online) Transverse momentum dependence of MCE scaled variance of negatively   
  charged particles ({\it left}) and the MCE correlation coefficient between positively   
  and negatively charged particles ({\it right}). Only particles in a mid-rapidity window   
  $|y|<0.3$ are measured.  Dashed lines denote the GCE result, Eq.(\ref{GCE_estimate}).}     
\label{StaticSource_dodpt_VI}   
\end{figure}   
  
In a small mid-rapidity window, with $|y|<0.3$, the effects of globally applied motional   
conservation laws cease  to be important (see Fig.(\ref{StaticSource_dodpt_VI})).   
Local correlations due to BE and FD statistics begin to dominate, and MCE deviations   
from the GCE results, Eq.(\ref{GCE_estimate}), are relevant only for the highest momentum bins.    
In BE or FD statistics we find for vanishing bin size~($\delta y,~\delta p_T$):    
\begin{equation}\label{GCE_estimate}   
\omega^{GCE}_{\delta} ~=~ \frac{\kappa_{2}^{N_{\delta},N_{\delta}}}{\kappa_{1}^{N_{\delta}}}  
~\simeq~ \frac{1}{1\pm e^{- \beta m_T \cosh y + \beta \mu}}~.  
\end{equation}   
BE and FD effects are strongest around mid-rapidity $y=0$. MCE calculations in   
Fig.(\ref{StaticSource_dodpt_VI}) are close to the GCE estimate Eq.(\ref{GCE_estimate}).   
In MB statistics we find only a weak $\Delta p_T$ dependence in a small mid-rapidity window.  
Please note that the acceptance scaling procedure predicts a Poisson distribution with  
$\omega \rightarrow 1$ and $\rho \rightarrow 0$ for all three statistics in the   
limit of very small acceptance.  
  
\subsection{Collectively Moving System}   
\label{BoostedSource}   

As established a long time ago, in order to properly define the thermodynamics
of a system with collective motion,  the partition function needs to be Lorentz 
invariant~\cite{Pathria_paper,Touschek}. The expectation values of observables need 
hence to transform according to the Lorentz transformation properties of these observables. 
In particular, the temperature $T$ is promoted to a four-vector $\beta^{\mu}$ (combining 
local temperature with collective velocity). The entropy, as well as particle multiplicities, 
remain Lorentz-scalars.

These requirements are in general not satisfied unless momentum conservation is put on an 
equal status with energy conservation. If the system is described by a MCE, 
then momentum should be conserved as well as energy \cite{Pathria_paper,Touschek}.
If the system is exchanging energy with a bath, the bath needs to exchange momentum as well. 

For ensemble averages, neglecting these rules and treating momentum differently from 
energy  
is safe as long as the system is 
close to the thermodynamic limit, since there ensembles become equivalent. The same is {\em not} 
true for fluctuation and correlation observables, which remain ensemble-specific \cite{SM_fluc_ce}.  

For a system at rest, these requirements are not apparent since the net momentum 
is zero. Statistical mechanics observables in a collectively moving system, however, lose their 
Lorentz invariance, if this is not maintained in the definition of the partition function.

To illustrate this point, we consider the properties of a system moving along the $z$-axis 
with a collective velocity given by Eq.(\ref{vec_u}). The total energy of the fireball is 
then $E=M \cosh(y_0)$, while its total momentum is given by $P_z=M \sinh(y_0)$. The mass 
of the fireball in its rest frame is $M=P^{\mu} u_{\mu}$. The system 4-temperature is 
$\vec \beta = \beta \vec u$. Local temperature and chemical potentials remain unchanged.   
We will use this section for a discussion of the second rank tensor   
(or co-variance matrix) $\kappa_2$, Eq.(\ref{kappa}).

The second order cumulant $\kappa_2$, Eq.(\ref{kappa}), is given by the second derivatives   
of the cumulant generating function with respect to the fugacities. Essentially this is the   
Hessian matrix \cite{hessian} of the function Eq.(\ref{Psi}), encoding the structure of its minima. The diagonal elements  
$\kappa_2^{X,X}$ are the variances of the GCE distributions of extensive quantities $X$.  
For example, $\kappa_2^{N_A,N_A}$ measures the GCE variance of the distribution of particle   
multiplicity of species $A$, while $\kappa_2^{Q,Q}$ denotes the GCE electric charge   
fluctuations, etc. The off-diagonal $ \kappa_2^{X,Y}$ elements give GCE co-variances   
of two extensive quantities $X$ and $Y$.    
  
For a boost along the $z$-axis the general co-variance matrix for a pion gas reads:     
\begin{align}   
\kappa_2 =    
\begin{pmatrix}   
\kappa_2^{N_A,N_A} & \kappa_2^{N_A,N_B} & \kappa_2^{N_A,Q} &   
\kappa_2^{N_A,E} & \kappa_2^{N_A,P_x} & \kappa_2^{N_A,P_y} & \kappa_2^{N_A,P_z} \\    
\kappa_2^{N_B,N_A} & \kappa_2^{N_B,N_B} & \kappa_2^{N_B,Q} &   
\kappa_2^{N_B,E} & \kappa_2^{N_B,P_x} & \kappa_2^{N_B,P_y} & \kappa_2^{N_B,P_z} \\    
\kappa_2^{Q,N_A} & \kappa_2^{Q,N_B} & \kappa_2^{Q,Q} & \kappa_2^{Q,E}    
& 0 & 0 &  0 \\    
\kappa_2^{E,N_A} & \kappa_2^{E,N_B} & \kappa_2^{E,Q} & \kappa_2^{E,E}    
& 0 & 0 & \kappa_2^{E,P_z}\\    
\kappa_2^{P_x,N_A} & \kappa_2^{P_x,N_B} & 0 & 0 & \kappa_2^{P_x,P_x} & 0 & 0 \\    
\kappa_2^{P_y,N_A} & \kappa_2^{P_y,N_B} & 0 & 0 & 0 & \kappa_2^{P_y,P_y} & 0 \\    
\kappa_2^{P_z,N_A} & \kappa_2^{P_z,N_B} & 0 & \kappa_2^{P_z,E} & 0 & 0 &  \kappa_2^{P_z,P_z}    
\end{pmatrix}~. \label{matrix}   
\end{align}   
Off-diagonal elements correlating a globally conserved charge with one of the momenta, i.e.   
$\kappa_2^{Q,P_x}$, as well as elements denoting correlations between different momenta, i.e.   
$\kappa_2^{P_x,P_y}$, are equal to zero due to antisymmetric momentum integrals.  
The values of elements correlating particle multiplicity and momenta, i.e.   
$\kappa_2^{N_A,P_x}$, depend strongly on the acceptance cuts applied. For fully phase   
space integrated ($4\pi$) multiplicity fluctuations and correlations these elements   
are equal to $0$, again due to antisymmetric momentum integrals.  
  
It is instructive to review the transformation properties of $\kappa$ under the Lorentz group:  
$\kappa_2^{X,Y}$ contains the correlations between 4-momenta $P^{\mu}$ and, in general, 
(scalar) conserved quantities and particle multiplicities $Q^j$. Hence, the elements 
$\kappa_2^{P^{\mu},P^{\nu}}$, i.e. $\ave{\Delta P^{\mu} \Delta P^{\nu}}$, will transform as a 
tensor of rank 2 under Lorentz transformations; $\kappa_2^{P^{\mu},Q^j }$, i.e. 
$ \ave{\Delta P^{\mu} \Delta Q^j }$, will transform as a vector (the rapidity distribution will 
simply shift); and the remaining $\kappa_2^{Q^j,Q^k }$ will be scalars.  
  
For a static system one finds for the co-variances $\kappa_2^{E,P_x}=\kappa_2^{E,P_y}  
=\kappa_2^{E,P_z}=0$.  Under these two conditions, a static system and full particle   
acceptance, the eigenvalues of the matrix Eq.(\ref{matrix}) factorise, and momentum   
conservation can be shown to drop out of the calculation \cite{acc}.   
  
For a boost along the $z$-axis (and arbitrary particle acceptance) it is the appearance of   
non-vanishing elements $\kappa_2^{P_z,E} = \kappa_2^{E,P_z} \not= 0$ which make the   
determinant of the matrix $\kappa_2$, Eq.(\ref{matrix}), invariant against such a boost.   
Please note that still $\kappa_2^{P_x,E} = \kappa_2^{P_y,E} = 0$.   
  
\begin{figure}[ht!]   
\epsfig{file=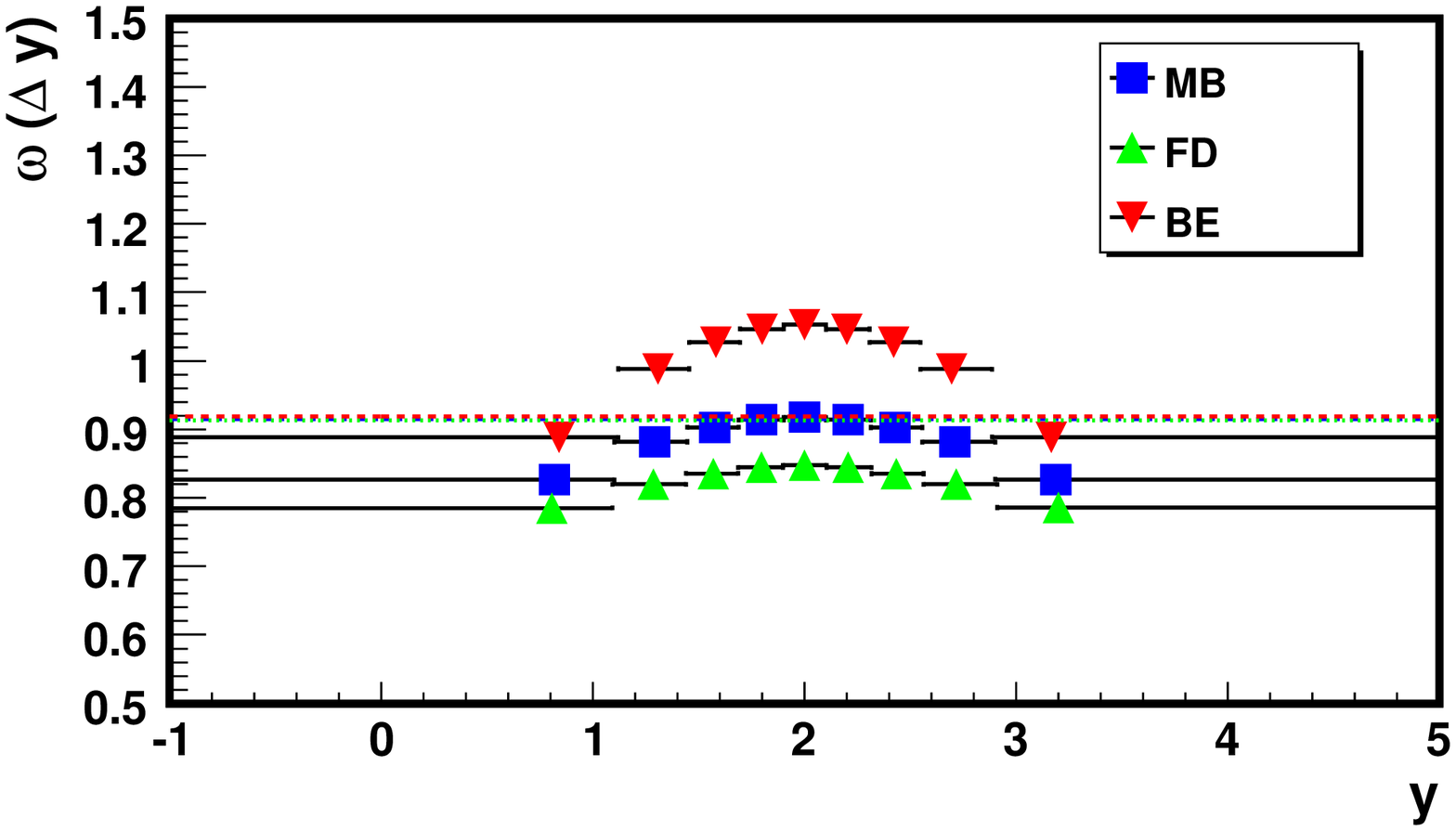,width=8.9cm,height=5.5cm}   
\epsfig{file=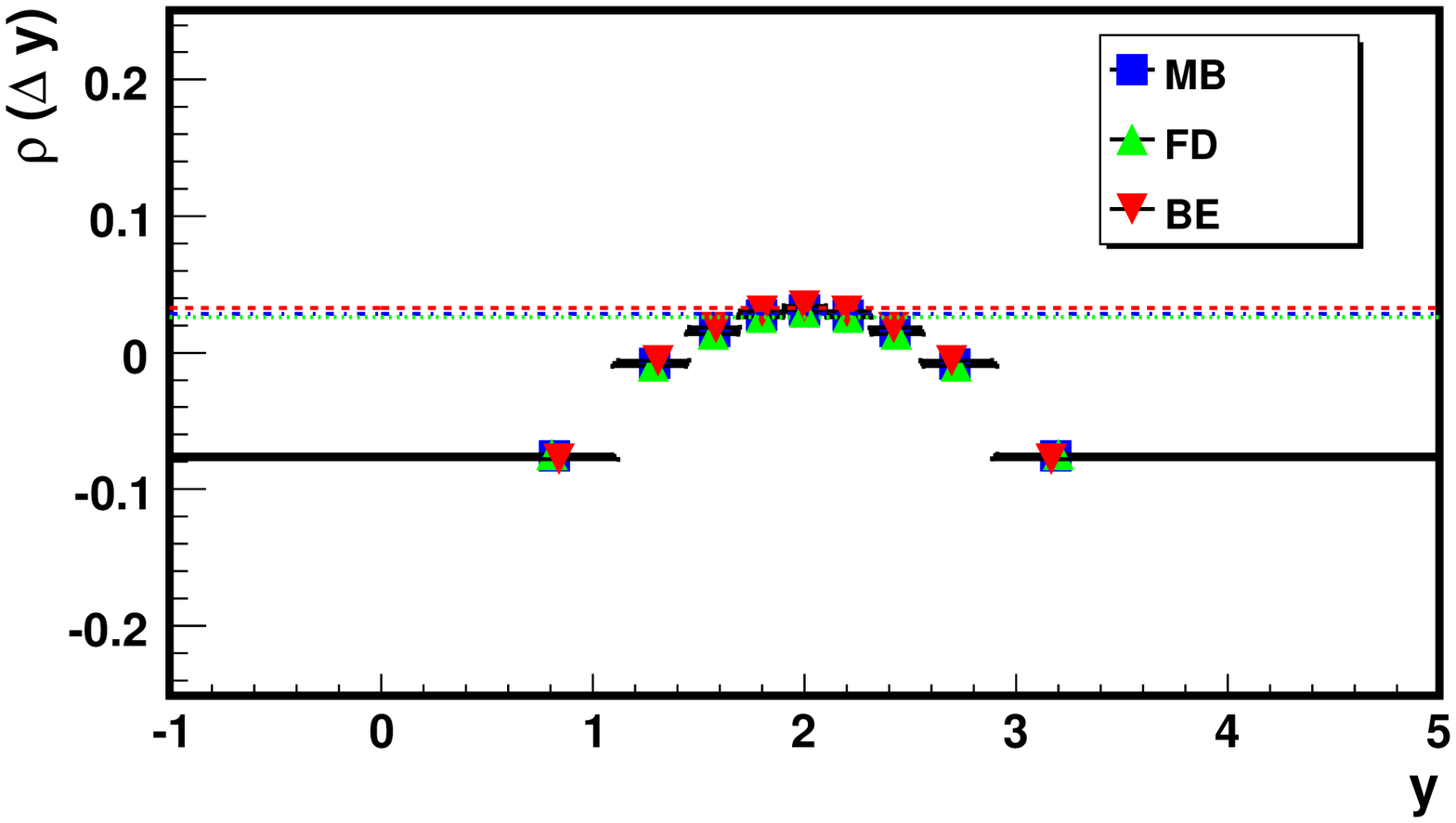,width=8.9cm,height=5.5cm}   
\caption{(Color online) Same as Fig.(\ref{StaticSource_dodp}) ({\it right}) and   
  Fig.(\ref{StaticSource_drhodp}) ({\it right}), but for a system moving with   
  $y_0 = 2$.}      
\label{MovingSource_dodp}   
\end{figure}   
In Fig.(\ref{MovingSource_dodp}) we show multiplicity fluctuations ({\it left})   
and correlations ({\it right}) for a system with boost $y_0=2$. The rapidity spectrum of   
Fig.(\ref{Boltzmannspectra}) ({\it right}) is simply shifted to the right by two units.   
The construction of  the acceptance bins is done as before. 
Multiplicity fluctuations and correlations within a bin transform as a 
vector (i.e., its $z$ component shifts in rapidity) as inferred from their Lorentz-transformation 
properties, provided both energy {\em and} momentum along the boost direction are conserved.      
  
This last point deserves attention because usually (starting from \cite{Fer50}) micro canonical 
calculations only conserve energy and {\em not} momentum.
Imposing exact conservation for energy, and only average conservation of momentum will 
make the system non-Lorentz invariant, since in a different frame from the co-moving 
one energy and momentum will mix, resulting in micro state-by-micro state fluctuations 
in {\em both} momentum and energy\footnote{This result is somewhat confusing, because energy-momentum is a vector of separately conserved currents.
It is therefore natural to assume that these currents can be treated within different ensembles; they are, after all, conserved separately.
It must be kept in mind, however, that it is not energy-momentum, but {\em particles} that are exchanged between the system and any canonical or grand canonical bath.  The amounts of energy and momentum carried by each particle are strictly correlated by the dispersion relation \cite{Touschek}.   In the situation examined here (unlike in a Cooper-Frye formalism \cite{cooperfrye}, where the system is ``frozen'' at the Freeze-out hypersurface, a {\em space-time} 4-vector correlated with 4-momentum) all {\em time dependence} within the system under consideration is absent due to the equilibrium assumption. Furthermore, the system is entirely thermal: the correlation between particle numbers when the system is sampled ``at different times'' is a $\delta-$function, that stays a $\delta-$function under all Lorentz-transformations.  Hence, unlike what happens in a Cooper-Frye freeze-out, energy-momentum and space-time do {\em not} mix in the partition function.  Together with the constraint from the particle dispersion relations, this means that different components of the 4-momentum need to be treated by the same ensemble, as explicitly demonstrated in this section.}.  
   
\begin{figure}[ht!]   
\epsfig{file=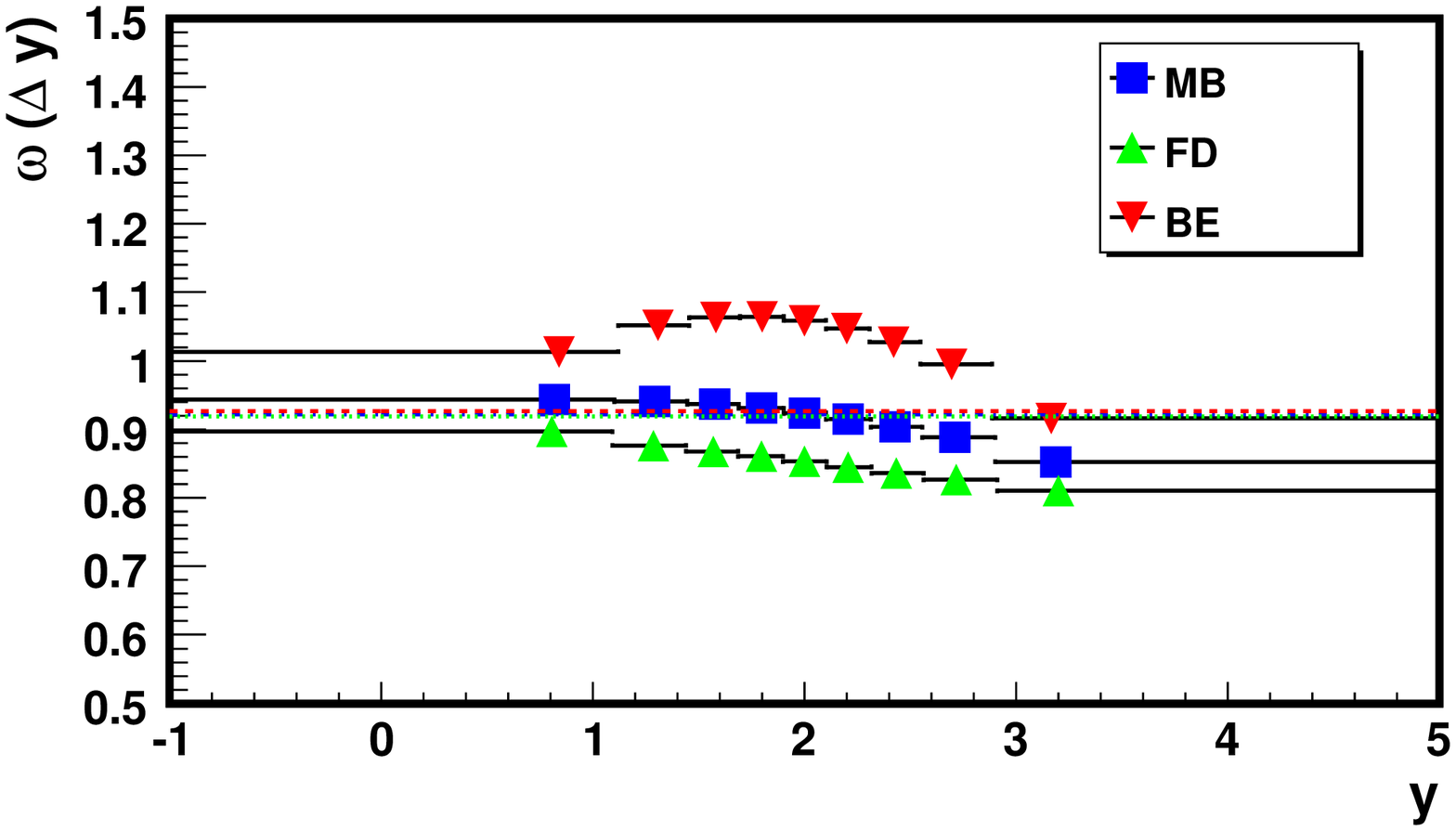,width=8.9cm,height=5.5cm}   
\epsfig{file=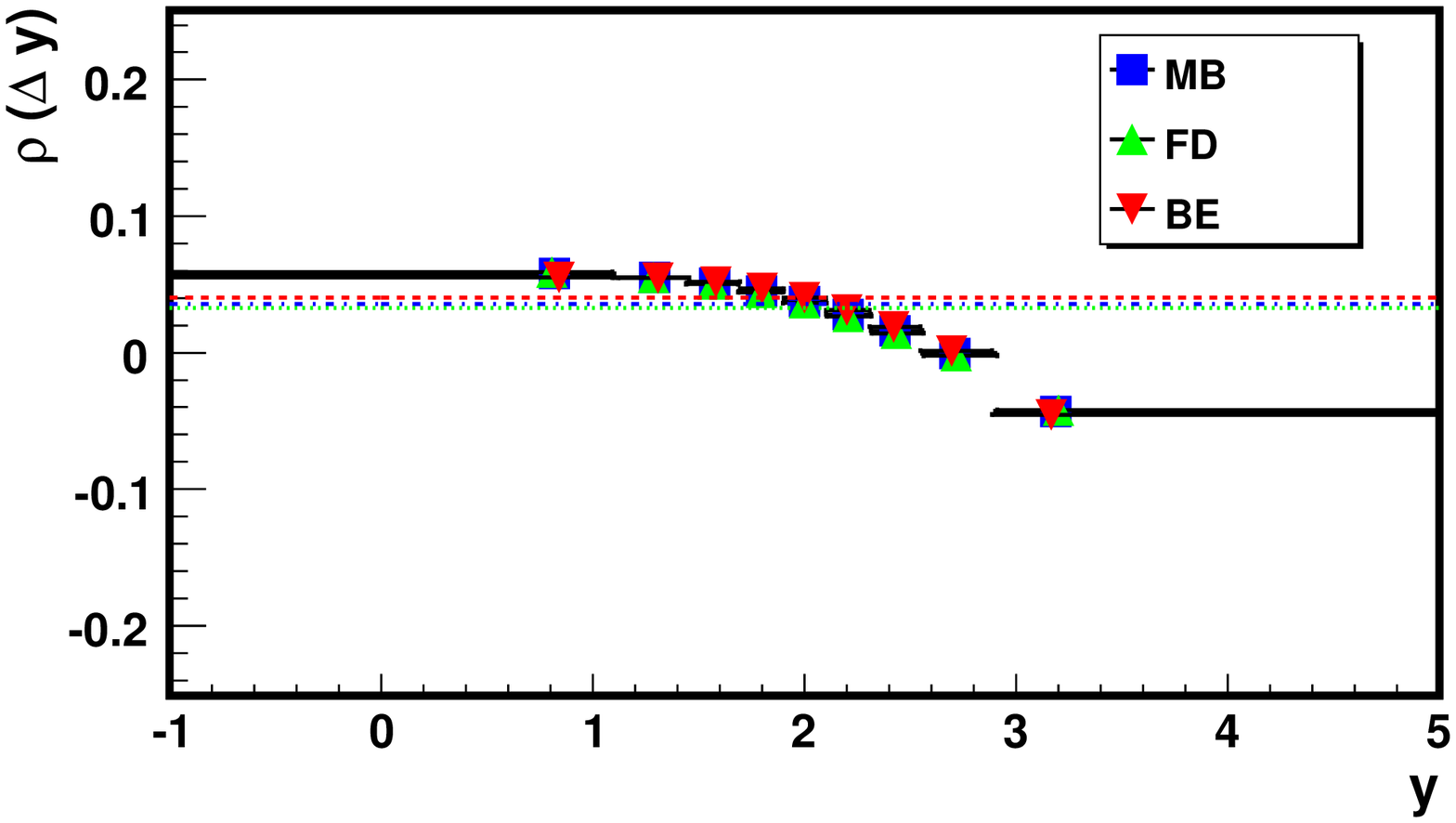,width=8.9cm,height=5.5cm}   
\caption{(Color online) Same as Fig.(\ref{MovingSource_dodp}), but without $P_z$ exact   
  conservation.}       
\label{MovingSource_dodp_nopz}   
\end{figure}  
This situation is explicitly shown in Fig.(\ref{MovingSource_dodp_nopz}).  Here, we have calculated 
multiplicity fluctuations and correlations in the same system as in Fig.(\ref{MovingSource_dodp}), 
but with exact conservation of {\em only} energy (and charge). In the co-moving frame of the 
system, the fluctuations and correlations are identical to Fig.(\ref{StaticSource_dxdy_nopz}). 
When the system is boosted, however, the distribution changes (not only by a shift in rapidity, 
as required by 
Lorentz-invariance), and loses its symmetry around the system's average boost.
  
This last effect can be understood from the fact that momentum does not have to be conserved 
event by event, but energy does.  It is easier, therefore, to create a particle with less rapidity 
than average (having less momentum than the boost, and parametrically less energy) than with more 
rapidity than average (having more momentum than the boost, and parametrically more energy) and 
still conserve energy overall. This leads to suppressed multiplicity fluctuations and a negative 
correlation coefficient for rapidity bins in the forward direction in comparison to rapidity bins 
in the backward direction. In Fig.(\ref{MovingSource_dodp}), where the system also needs to 
conserve momentum exactly, this enhancement is balanced by the fact that it will be more difficult 
to conserve momentum when particles having less momentum than the boost are created.  
 
A situation such as that in Fig.(\ref{MovingSource_dodp_nopz}) is impossible to be realized 
physically.  It {\em could}, however, be realized within ``system in a box''-type calculations 
with non-equilibrium models: E.g., a transport model inside an infinitely heavy box (that absorbs 
momentum but not energy event-by-event) would end up exhibiting micro canonical correlations 
similar to those in Fig.(\ref{MovingSource_dodp_nopz}). A similar box with `periodic` walls, 
however, would conserve energy as well as momentum inside the box, and should therefore behave 
as in Fig.(\ref{MovingSource_dodp}). Thus, correlations within boosted sources provide a sensitive 
test of the Lorentz-invariance of such transport models.

\section{Correlations between Bins disconnected in Momentum Space}   
\label{LongRangeCorr}   
``Long range correlations'' between bins well disconnected in momentum have been suggested   
to arise from dynamical processes. Examples include color glass   
condensate~\cite{correlation_cgc1,correlation_cgc2}, droplet formation driven hadronization   
\cite{ourbulk1}, and phase transitions within a percolation-type mechanism \cite{phasetrans_lrc,cp_lrc}.  
The elliptic flow measurements, widely believed to signify the production of a liquid at   
RHIC \cite{whitebrahms,whitephobos,whitestar,whitephenix}, are also, ultimately,   
correlations between particles disconnected in phase space (here, the azimuthal angle).  
  
As we will show, however, conservation laws will also introduce such correlations between   
any two  (connected or not) distinct regions of momentum space. No dynamical effects are   
taken into consideration (only an isotropic thermal system).  
  
\begin{figure}[ht!]   
\epsfig{file=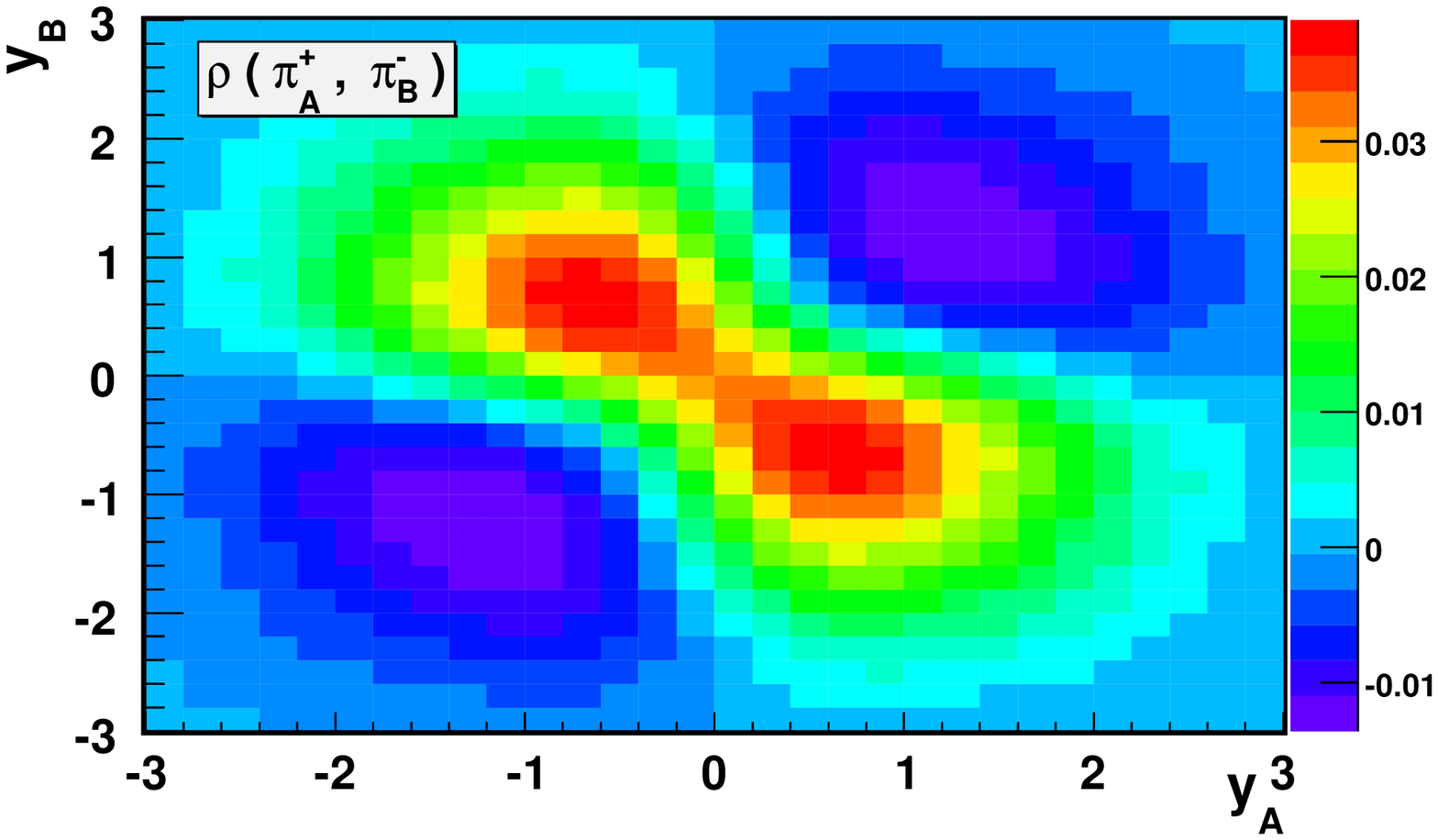,width=8.9cm,height=5.5cm}   
\epsfig{file=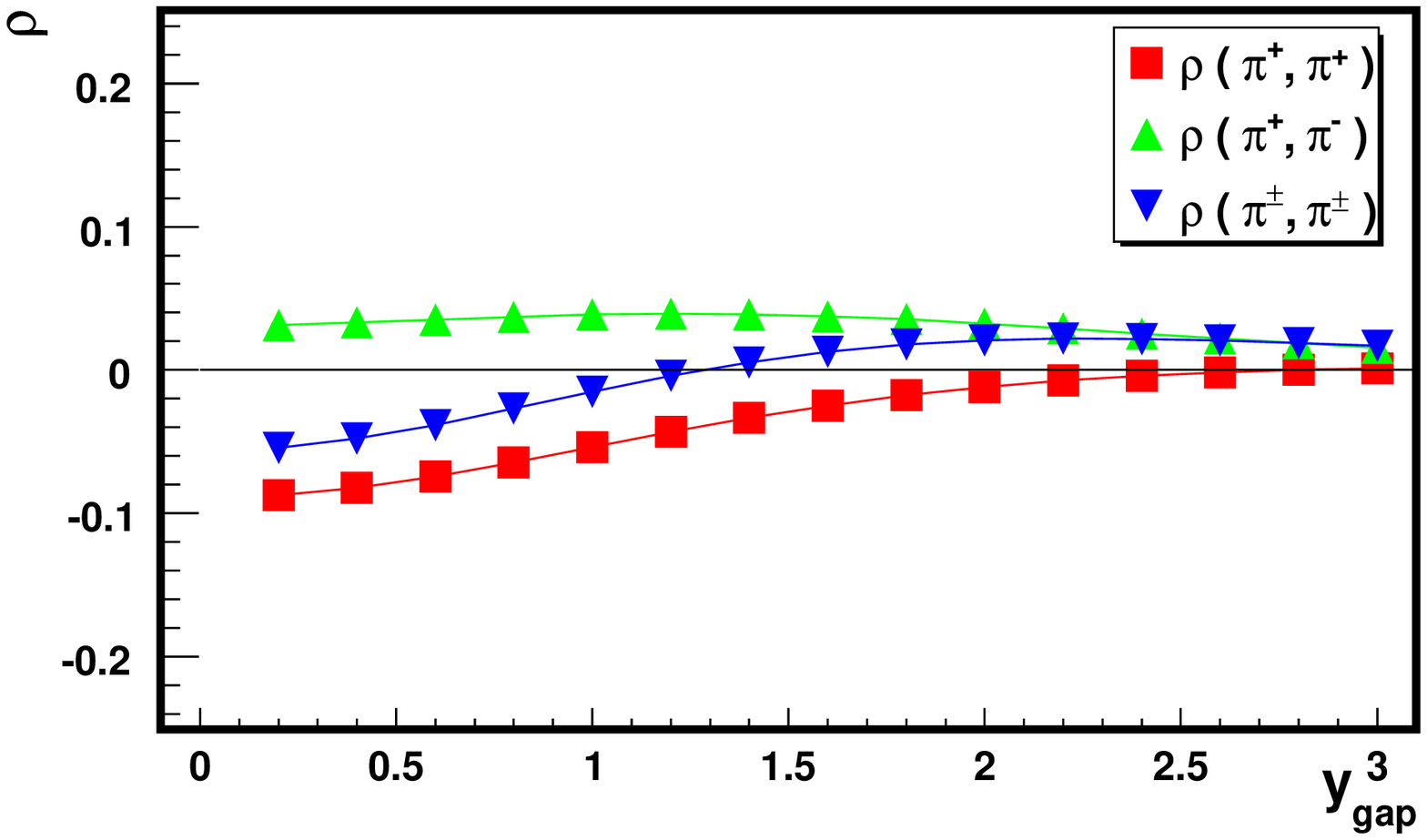,width=8.9cm,height=5.5cm}   
\caption{(Color online) {\it Left}: The correlation coefficient $\rho$ between the multiplicity of   
positively charged particles in a bin located at $y_A$ with negatively charged particles   
in a bin located at  $y_B$, both with a 0.2 width in rapidity.   
{\it Right}: The correlation coefficient between multiplicities in two bins separated by $y_{gap}$   
 of like, unlike, and all charged particles. Both plots show MCE MB results.}  
\label{FB_corr}  
\end{figure}

Let us first consider correlations between the multiplicities of particles $A$ and $B$,   
within two bins centered around $y_A$ and $y_B$, with (constant) widths   
$\Delta y_A= \Delta y_B =0.2$. In Fig.(\ref{FB_corr}) ({\it left}) we show the correlation   
coefficient, calculated using Eq.(\ref{rho}), between positive and negative particles as   
a function of $y_A$ and $y_B$. In Fig.(\ref{FB_corr}) ({\it right}) we show the correlation  
 coefficient between like-charge, unlike-charge, and all charged particles as a function   
of $y_{gap} = y_A - y_B.$  
  
Energy conservation always leads to anti-correlation between different  
momentum space bins. Charge conservation leads to a positive correlation of unlike charged   
particles and anti-correlation of like-sign particles. Longitudinal momentum conservation,   
however, is responsible for the structure visible in Fig.(\ref{FB_corr})({\it left}).   
Having a small (large) number of particles in a bin with   
positive average longitudinal momentum, leads to a larger (smaller) number of   
particles in a bin with different but also positive $P_z$, (blue dips).   
This makes also a state with smaller (larger) particle number with   
opposite average longitudinal momentum $-P_z$ more likely (red hills).  
At large values of $y_A$ the correlation coefficient $\rho \simeq 0$ for any $y_B$, because   
the yield $\langle N_A \rangle$ in $\Delta y_A$ is asymptotically vanishing.   
  
In Fig.(\ref{FB_corr}) ({\it right}) we show the correlation coefficient   
along the diagonal from top left to bottom right as a function of separation. Unlike-sign   
particles are positively correlated. Like-sign and all charged   
particles are negatively correlated  at small separation $y_{gap}$.   
For large separation the correlation becomes asymptotically zero, because the yield   
is zero. However, please note that in particular   
$\rho (\pi^{\pm},\pi^{\pm}) > \rho (\pi^{+},\pi^{-})$ at large $y_{gap}$.  
 $P_z$ conservation is dominant.   
  
Disregarding $P_z$ conservation would destroy the particular structure in   
Fig.(\ref{FB_corr}) ({\it left}) and lead to a single peak at the origin.   
The correlation would then be insensitive to the momentum direction, and only be sensitive   
to the energy content of a bin $\Delta y$. The observables in Fig.(\ref{FB_corr})   
transform under boosts ($y_{A,B} \rightarrow y_{A,B}-y_{0}$), provided momentum  
along the boost axis is exactly conserved.  
  
\begin{figure}[ht!]  
\epsfig{file=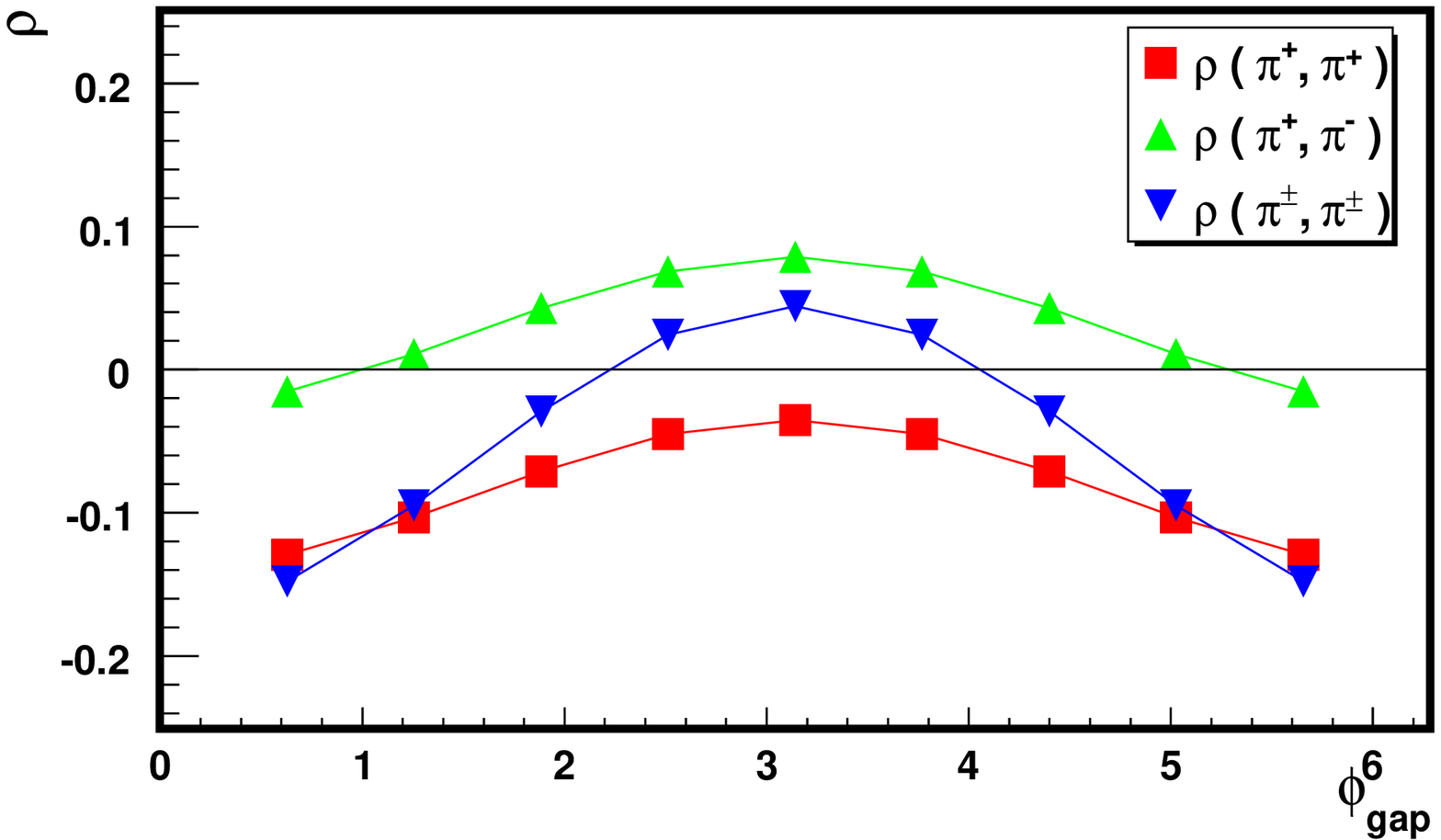,width=8.9cm}  
\epsfig{file=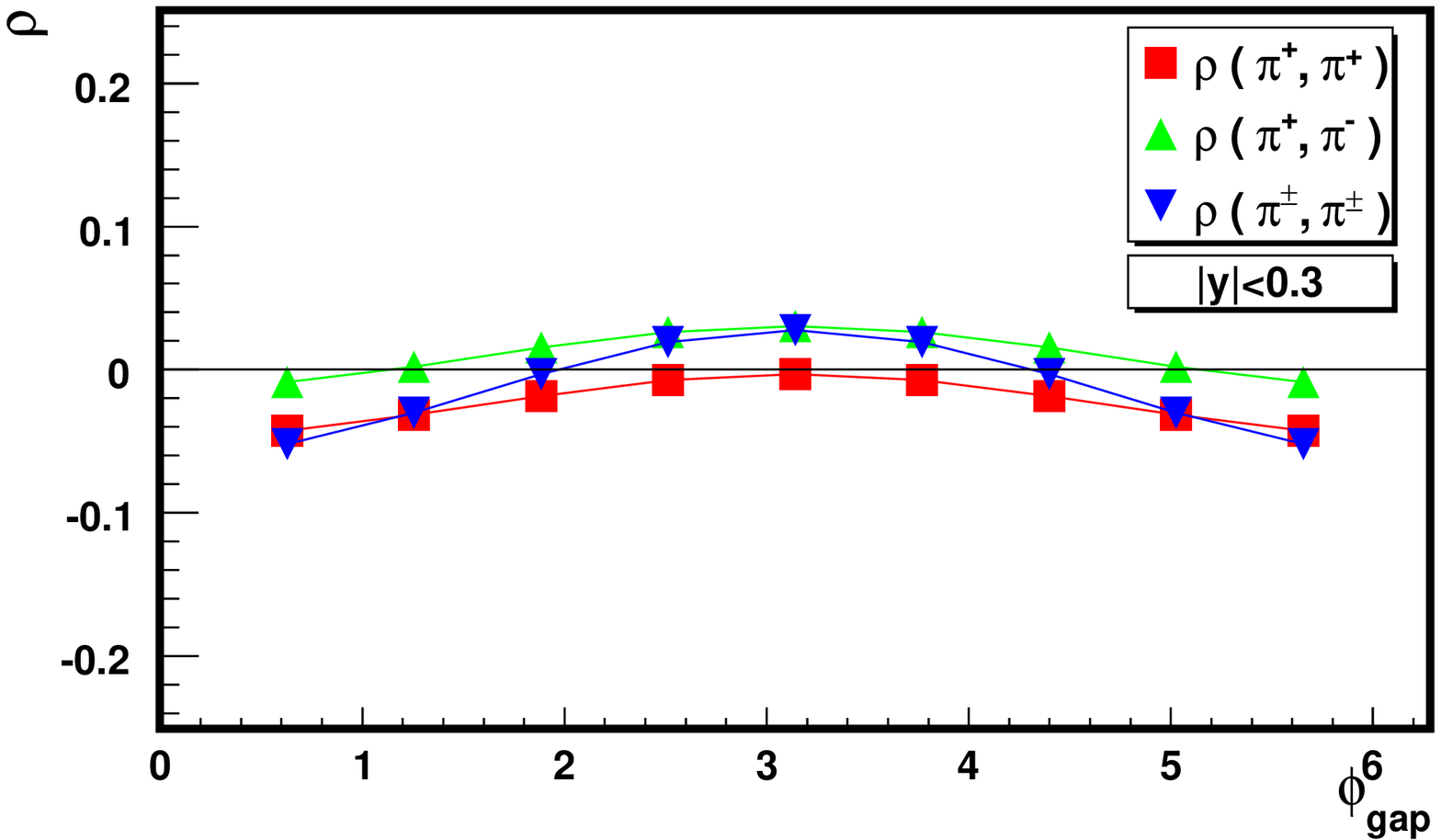,width=8.9cm}  
\caption{(Color online) The correlation coefficients of particles in distinct $\Delta \phi$ bins as a   
function of separation $\phi_{gap}$ in azimuth. ({\it left}) integrated over all   
phase space. ({\it right}) only particles with $|y|<0.3$ are observed.   
Both plots show MCE MB results. No elliptic flow is considered.}    
\label{Angular_corr}  
\end{figure}

Angular correlations also arise due to conservation of transverse momenta $P_x$   
and $P_y$. In Fig.(\ref{Angular_corr}) we show the correlation coefficient   
 between particles in different $\Delta \phi$ bins. The flat\footnote{Since we consider globally   
equilibrated systems, elliptic flow is disregarded here.} angular spectrum $dN / d\phi$  
has been divided into 10 equal size bins and the correlation coefficient is presented   
as a function of separation of the centers of the corresponding bins.   
  
To explain Fig.(\ref{Angular_corr}) we note firstly that when disregarding  
 exact conservation of $P_x$ and $P_y$ the correlation coefficients are insensitive   
to the distance $\phi_{gap}$ of any two bins. Only the correlations due to energy and   
charge conservation affect the result. Charge conservation leads to correlation   
of unlike-sign particles and to anti-correlation of like-sign particles. Energy   
conservation always anti-correlates multiplicities in two bins. For   
$\rho \left(\pi^{\pm},\pi^{\pm} \right)$ the effect of charge conservation cancels for a   
neutral system, however, effects of energy-momentum conservation are stronger,   
as a larger number of particles (hence a larger part of the total system) is observed.  
  
Conservation of transverse momenta $P_x$ and $P_y$ is now responsible for the   
$\phi_{gap}$ dependence of $\rho$. The line of arguments is similar to the ones   
before: Observing a larger (smaller) number of particles in some bin at $\phi_0$  
 implies that, in order to balance momenta $P_x = P_y = 0$, one should also observe a   
larger (smaller) number of particles in the opposite direction $\pi-\phi_0$. A larger (smaller)   
number of particles in a bin with $\phi_{gap} = \pi/2$ would do little to help   
to balance momentum, but conflict with energy conservation.

\section{Discussion and Summary}      
\label{Summary}   
We have presented multiplicity fluctuations and correlations in limited momentum bins   
for ideal relativistic gases in the MCE in the thermodynamic limit.   
For our examples we chose a gas with three degenerate massive particles   
(positive, negative, neutral) in three different statistics (Maxwell-Boltzmann,   
Fermi-Dirac, Bose-Einstein).  
  
For the width of multiplicity distributions in limited bins of momentum space  
a simple and intuitive picture emerges. In the Maxwell-Boltzmann approximation one finds   
a wider distribution for momentum bins with low average momentum when compared to bins  
with higher average momentum, but same average particle number.   
This qualitative behaviour is a direct consequence of energy and momentum conservation.   
The results in Fermi-Dirac and Bose-Einstein statistics, furthermore, show pronounced   
effects at the low momentum tail of the spectrum.  
  
The correlation coefficient additionally shows a similar qualitative behaviour. In bins   
with low average momentum the correlation coefficient between positively and negatively   
charged particles is indeed positive, as one would expect from charge conservation. However,  
in bins with large average momentum the effects of joint energy and momentum conservation  
can lead to anti-correlated distributions of unlike-charged particles.   
  
For boosted systems we found that the role of exactly imposed motional   
conservation laws is particularly important. Fluctuations and correlations transform   
 under boosts, provided momentum conservation along the boost direction is taken   
into account. This ensures, in particular, that they become boost invariant if the   
underlying system is boost-invariant.  
  
Lastly, we found that even in the thermodynamic limit long range correlations between   
disconnected regions in momentum space prevail. Multiplicities in different   
rapidity bins, as well as different bins in azimuth, have a non-zero correlation   
coefficient.   
  
It is premature to use the model presented here for a {\em quantitative} comparison with experimental data. Firstly, the inclusion of resonances will provide important corrections.   These can be implemented within our model using Monte-Carlo techniques, and will be the subject of a subsequent work \cite{mc}.   

An additional effect missing here that could change results qualitatively is longitudinal flow.  It can be seen, ``by symmetry'', that correlations and fluctuations in a perfectly boost-invariant fluid would be, just like other physical observables, independent of rapidity.  Reconciling this with the calculations in sections \ref{StaticSource} and \ref{BoostedSource}  would require calculating correlations of many independent sources, each centered around a particular rapidity.

It is however {\em not} currently clear whether experimental data, even at highest RHIC and LHC energies at mid-rapidity, approximates this limit.   At lower SPS energies, measurements ~\cite{BeniData} of the rapidity and transverse momentum dependence of particle multiplicity fluctuations show qualitatively similar results to those of our calculations (it should be noted that, as shown in \cite{benchmark}, this behaviour for fluctuations also arises in molecular dynamics models, where conservation laws are included but equilibrium is not assumed). It has long been noticed \cite{busza,busza2} that many observables binned in rapidity obey ``universal fragmentation'', suggesting that a ``Landau hydrodynamics'', with negligible initial longitudinal flow even away from mid-rapidity, might be more appropriate than boost-invariance to describe the initial state, even at ultra-relativistic energies \cite{steinberg}.   Experimentally, these measurements might be non-trivial since the size of the bins must be large enough for conservation laws to have an effect, but, since RHIC has large-acceptance data \cite{whitephobos,whitebrahms} and the LHC is planning a larger-acceptance detector \cite{totem}, they are in principle possible.

If boost-invariance is not really there, then correlations and fluctuations binned by rapidity should be qualitatively similar to those calculated in section \ref{StaticSource} even at RHIC and LHC energies.  We therefore suggest the experimental measurement of such fluctuations (Figs.(\ref{StaticSource_dodp},\ref{StaticSource_drhodp}) ({\it right})) at these energies as an experimental probe of the degree of boost-invariance of the system.  

Similarly, transverse flow should not change the correlation and fluctuations within limited transverse momentum bins (the left panels of Figs.(\ref{StaticSource_dodp},\ref{StaticSource_drhodp})) beyond a trivial shift {\em provided} there are no significant fluctuations in collective flow observables.  These fluctuations are widely expected to arise in an imperfect fluid \cite{vogel}, but have remarkably not been observed, for example, in elliptic flow measurements \cite{loizides,sorensen}.  We suggest, therefore, that a measurement of $p_T$ binned fluctuations and correlations could provide a qualitative way to assess the magnitude of event-by-event flow observables wrt to the thermal observables presented here.

Finally, effects of the sort studied in section \ref{LongRangeCorr} will surely appear in any measurement looking for correlations across momentum space.
Long range rapidity correlations have, in particular, been advocated as a signature of new physics \cite{correlation_cgc1,correlation_cgc2}.   Our work shows that correlations due to conservation laws actually cover as wide a range in rapidity as those measured in \cite{tarnowsky,PHENIX_rap_corr,PHOBOS_rap_corr}.  The magnitude of the correlations, however, is significantly (as much as an order of magnitude) lower than either the experimental result or any reasonable ``new physics'', suggesting that the origin of the experimentally observed correlations lies elsewhere. In particular, correlations induced by initial state geometry are considerably larger than those induced by conservation laws, as a comparison between Fig.(\ref{FB_corr}) and the results in \cite{GMC} will show\footnote{Note, however, that these results were obtained in the infinite volume limit.  Finite volume effects are likely to increase the strength of correlations arising from conservation laws, though for realistic nuclear volumes such corrections should not alter the results by an order of magnitude}.  Nevertheless, energy-momentum conservation does trigger correlations across a wide rapidity interval, and, as shown in \cite{benchmark}, qualitatively the magnitude of these correlations is independent of the degree of equilibration of the system, so their presence in experimental data is very plausible.   Perhaps complementing the rapidity correlations with azimuthal correlation measurements, such as those in Fig.(\ref{Angular_corr}), might  clarify their role, although the latter are particularly susceptible to ``non-trivial'' physics contributions, such as jet pairs and elliptic flow. 

In conclusion, we have presented microcanonical ensemble calculations of correlation and fluctuation observables within and across bins within a range of rapidity and transverse momenta. The calculations presented here provide {\em qualitative} effects   
affecting  multiplicity fluctuations and correlations. These effects arise solely from   
statistical mechanics and conservation laws. It will be extremely important to see whether   
these qualitative effects are visible in further experimental measurements of the momentum   
dependence of multiplicity fluctuations and correlations. If so, these effects might well   
be of similar magnitude to the signals for new physics. Disentangling them from   
dynamical correlations will then be an important, and likely non-trivial task.   
   
\begin{acknowledgments}    
We would like to thank  F.~Becattini, M.~Bleicher,  
E.L.~Bratkovskaya, W.~Broniowski, J.~Cleymans, L.~Ferroni, M.I.~Gorenstein, M.~Ga\'zdzicki,  
S.~H\"aussler, V.P.~Konchakovski, B.~Lungwitz, S.~Jeon and J.~Rafelski for fruitful  
discussions.   
G.T. acknowledges the financial support received from the Helmholtz International  
Center for FAIR within the framework of the LOEWE program  
(Landesoffensive zur Entwicklung Wissenschaftlich-\"Okonomischer  
Exzellenz) launched by the State of Hesse.  
\end{acknowledgments}    
   
\appendix   

\section{MCE Partition Function}   
\label{MCEPF}  
This section serves to provide a connection between Eqs.(\ref{P_one}-\ref{P_three})   
and Eq.(\ref{Curly}); namely to prove the following relation:  
\begin{equation}\label{Connection}  
\mathcal{Z}^{\vec Q,\vec P}(V,\beta,\vec \mu,\vec u) =   
e^{- P^{\mu} u_{\mu} \beta} ~ e^{Q^j \mu_j \beta} ~ Z_{MCE}(V,\vec P,\vec Q)~,  
\end{equation}  
where $Z_{MCE}(V,\vec P,\vec Q)$ is the standard MCE partition function for  
a system of volume $V$, collective four-momentum $\vec P$ and a set of conserved   
Abelian charges $\vec Q$, as worked out in \cite{fourtemp,fourtempII}. The MCE partition function  
$Z_{MCE}(V,\vec P,\vec Q)$ counts the number of micro states consistent with this   
set of fixed extensive quantities. Likewise one could interpret the number  
$\mathcal{Z}^{\vec Q,\vec P}(V,\beta,\vec \mu,\vec u)$ as the number of micro states   
with the same set of extensive quantities for a GCE with local inverse temperature $\beta$,  
four-velocity $\vec u$, and chemical potentials $\vec \mu$.  
  
The starting point for this calculation is our Eq.(\ref{Curly}):  
\begin{equation}\label{Curly_app}  
\mathcal{Z}^{\vec Q,\vec P}(V,\beta,\vec \mu,\vec u) ~=~   
\int \limits_{-\pi}^{\pi} \frac{d^J\phi}{\left( 2\pi \right)^J}   
~e^{-iQ^j \phi_{j}}    
~\int \limits_{-\infty}^{\infty} \frac{d^4\alpha }{\left( 2\pi \right)^4}    
~ e^{-iP^{\mu} \alpha_{\mu}}   
~\exp \Bigg[V \Psi \left(\beta,\vec \mu,\vec u ; \vec \phi, \vec \alpha \right) \Bigg]~.  
\end{equation}  
Let us take a closer look at the exponential of Eq.(\ref{Curly_app}). For this we  
spell out Eq.(\ref{Psi}) and use the substitutions Eqs.(\ref{subst_1})   
and (\ref{subst_2}):  
\begin{equation}  
\exp \left[ \sum_l \frac{Vg_l}{\left(2 \pi \right)^3} \int d^3p \ln \left( 1 \pm   
e^{-p_l^{\mu}\left(\beta u_{\mu} - i\alpha_{\mu} \right)}   
e^{q_l^j \left(\beta \mu_j + i\phi_j \right)}   
\right)^{\pm 1} \right]~.  
\end{equation}  
Expanding the logarithm yields:  
\begin{equation}\label{Psi_app}  
\exp \left[ \sum_l \frac{Vg_l}{\left(2 \pi \right)^3} \int d^3p   
\sum \limits_{n_l=1}^{\infty}  \frac{\left(\mp 1 \right)^{n_l}}{n_l}  
e^{-n_l~\! p_l^{\mu}\left(\beta u_{\mu} - i\alpha_{\mu} \right)}   
e^{n_l~\! q_l^j \left(\beta \mu_j + i\phi_j \right)}   
 \right]~.  
\end{equation}  
Replacing now the momentum integration in Eq.(\ref{Psi_app}) by the usual summation over  
 individual momentum levels $\frac{V}{\left(2 \pi \right)^3} \int d^3p \rightarrow \sum_{k_{n_l}}$   
gives:  
\begin{equation}  
\exp \left[ \sum_l \sum \limits_{n_l=1}^{\infty} \sum \limits_{k_{n_l}}   
 \frac{g_l \left(\mp 1 \right)^{n_l}}{n_l}  
e^{-n_l~\! p_{k_{n_l}}^{\mu}\left(\beta u_{\mu} - i\alpha_{\mu} \right)}   
e^{n_l~\! q_l^j \left(\beta \mu_j + i\phi_j \right)}   
 \right]~.  
\end{equation}  
Finally, expanding the exponential yields:  
\begin{eqnarray}  
\mathcal{Z}^{\vec Q,\vec P}(V,\beta,\vec \mu,\vec u) &=&   
\int \limits_{-\pi}^{\pi} \frac{d^J\phi}{\left( 2\pi \right)^J}   
~e^{-iQ^j \phi_{j}}    
~\int \limits_{-\infty}^{\infty} \frac{d^4\alpha }{\left( 2\pi \right)^4}    
~ e^{-iP^{\mu} \alpha_{\mu}}   
~\prod_l \prod \limits_{n_l=1}^{\infty} \prod \limits_{k_{n_l}} \sum_{c_{k_{n_l}} =0}^{\infty}  \nonumber \\  
 && \frac{1}{c_{k_{n_l}}!}  
\left( \frac{g_l \left(\mp 1 \right)^{n_l}}{n_l} \right)^{c_{k_{n_l}}}  
e^{-c_{k_{n_l}} n_l~\! p_{k_{n_l}}^{\mu}\left(\beta u_{\mu} - i\alpha_{\mu} \right)}   
e^{c_{k_{n_l}} n_l ~\! q_l^j \left(\beta \mu_j + i\phi_j \right)}~.  
\end{eqnarray}  
Only sets of numbers $\{c_{k_{n_l}}\}$ which meet the requirements:  
\begin{equation}  
\sum_l \sum \limits_{n_l=1}^{\infty} \sum \limits_{k_{n_l}}   
c_{k_{n_l}} n_l ~ p_{k_{n_l}}^{\mu} = P^{\mu}   
\qquad  \textrm{and} \qquad  
\sum_l \sum \limits_{n_l=1}^{\infty} \sum \limits_{k_{n_l}}  
c_{k_{n_l}} n_l ~ q_l^j = Q^j~,  
\end{equation}  
have a non-vanishing contribution to the integrals. Therefore we can pull   
these factors in front of the integral:  
\begin{eqnarray}  
\mathcal{Z}^{\vec Q,\vec P}(V,\beta,\vec \mu,\vec u) &=&   
e^{- P^{\mu} u_{\mu} \beta} ~ e^{Q^j \mu_j \beta}   
\int \limits_{-\pi}^{\pi} \frac{d^J\phi}{\left( 2\pi \right)^J}   
~e^{-iQ^j \phi_{j}}    
~\int \limits_{-\infty}^{\infty} \frac{d^4\alpha }{\left( 2\pi \right)^4}    
~ e^{-iP^{\mu} \alpha_{\mu}}  \nonumber \\  
 &&\prod_l \prod \limits_{n_l=1}^{\infty} \prod \limits_{k_{n_l}} \sum_{c_{k_{n_l}} =0}^{\infty}  
\frac{1}{c_{k_{n_l}}!} \left( \frac{g_l \left(\mp 1 \right)^{n_l}}{n_l} \right)^{c_{k_{n_l}}}  
e^{ i c_{k_{n_l}} n_l ~\! p_{k_{n_l}}^{\mu}  \alpha_{\mu} }   
e^{ i c_{k_{n_l}} n_l ~\! q_l^j \phi_j } ~.  
\end{eqnarray}  
Reverting the above expansions one returns to the definition of $Z_{MCE}(V,\vec P,\vec Q)$   
from Ref.\cite{fourtemp,fourtempII} times the Boltzmann factors:  
\begin{eqnarray}  
\mathcal{Z}^{\vec Q,\vec P}(V,\beta,\vec \mu,\vec u) &=&   
e^{- P^{\mu} u_{\mu} \beta} ~ e^{Q^j \mu_j \beta}   
\int \limits_{-\pi}^{\pi} \frac{d^J\phi}{\left( 2\pi \right)^J}   
~e^{-iQ^j \phi_{j}}    
~\int \limits_{-\infty}^{\infty} \frac{d^4\alpha }{\left( 2\pi \right)^4}    
~ e^{-iP^{\mu} \alpha_{\mu}}  \nonumber \\  
&& \exp \left[ \sum_l \frac{Vg_l}{\left(2 \pi \right)^3} \int d^3p \ln \left( 1 \pm   
e^{ i p_l^{\mu}\ \alpha_{\mu} }~ e^{ i q_l^j \phi_j }   
\right)^{\pm 1} \right]~,  
\end{eqnarray}  
which proves Eq.(\ref{Connection}). Therefore we write for the   
GCE distribution of extensive quantities:  
\begin{equation}   
P_{gce}(\vec Q, \vec P) ~=~ \frac{e^{- P^{\mu} u_{\mu} \beta} ~ e^{Q^j \mu_j \beta}~    
Z_{MCE}(V,\vec P,\vec Q)}{Z_{GCE}(V,\beta,\vec \mu, \vec u)} ~=~   
\frac{\mathcal{Z}^{\vec Q,\vec P}(V,\beta,\vec \mu,\vec u)}{Z_{GCE}(V,\beta,\vec \mu, \vec u)}~,  
\end{equation}  
which provides the promised connection between Eqs.(\ref{P_one}-\ref{P_three})   
and Eq.(\ref{Curly}).

\section{Asymptotic Joint Distribution}   
\label{AJD}  
The MCE joint multiplicity distribution $P_{mce}(N_A,N_B)$ is   
conveniently expressed by the ratio of two GCE joint distributions:   
\begin{eqnarray}   
P_{mce}(N_A,N_B) &=& P_{gce}(N_A,N_B|B,S,Q,E,\dots)~, \\   
&=&\frac{P_{gce}(N_A,N_B,B,S,Q,E,\dots)}{P_{gce}(B,S,Q,E,\dots)}~.   
\end{eqnarray}   
In the TL the distributions $P_{gce}(N_A,N_B,B,S,Q,E,\dots)$ and $P_{gce}(B,S,Q,E,\dots)$   
can be approximated by MND's, Eq.(\ref{MND}). The charge vector Eq.(\ref{DQ_k}) for a MCE HRG   
with three conserved charges would read:  
\begin{equation}   
\left( \Delta \mathcal{Q} \right) ~=~ \left(\Delta N_A,~ \Delta N_B,~ \Delta B,~ \Delta   
  S,~ \Delta Q,~ \Delta E,~\dots\right)~.  
\end{equation}   
Evaluating the MND, Eq.(\ref{MND}), around its peak for ($B,S,Q,E,\dots$) yields:  
\begin{equation}   
\left( \Delta \mathcal{Q} \right)  
~=~ \left(\Delta N_A,~ \Delta N_B, ~0,~0,~ 0,~0,~\dots\right)~.  
\end{equation}   
The vector Eq.(\ref{xi}) then becomes:   
\begin{align}   
\vec \xi  
~=~ V^{-1/2} ~   
\begin{pmatrix}   
\lambda_{1,1} ~\Delta N_A + \lambda_{1,2} ~ \Delta N_B \\    
\lambda_{2,1} ~\Delta N_A + \lambda_{2,2} ~ \Delta N_B \\    
\lambda_{3,1} ~\Delta N_A + \lambda_{3,2} ~ \Delta N_B \\    
\lambda_{4,1} ~\Delta N_A + \lambda_{4,2} ~ \Delta N_B \\    
\lambda_{5,1} ~\Delta N_A + \lambda_{5,2} ~ \Delta N_B \\    
\dots \\    
\end{pmatrix}~,   
\end{align}   
where $\lambda_{i,j} $ are the elements of the matrix Eq.(\ref{invsigma}).   
Therefore:  
\begin{equation}\label{xijxij}  
\xi_j ~ \xi^j  
~=~ V^{-1}~\Big[ \left( \Delta   
  N_A \right)^2 \sum \limits_{j=1}^J ~    
\lambda_{j,1}^2 ~+~2\left( \Delta N_A \right) \left( \Delta N_B \right) \sum   
\limits_{j=1}^J ~     
\lambda_{j,1} \lambda_{j,2}   ~+~ \left( \Delta N_B \right)^2 \sum   
\limits_{j=1}^J ~ \lambda_{j,2}^2 \Big]~,  
\end{equation}   
with $J=2+3+4=9$ for a MCE HRG with momentum conservation.   
Using Eq.(\ref{xijxij}), the micro canonical joint multiplicity distribution of   
particle species $A$ and $B$ can thus be written as:  
\begin{equation}\label{PCE}   
 P_{mce}(N_A,N_B) ~=~  \frac{1}{\left(2\pi V\right)} ~\frac{\det   
  \sigma_N}{\det \sigma} ~\exp \left[-\frac{1}{2}~   
\xi_j~ \xi^j \right]~,  
\end{equation}   
where $ \sigma_N$ is the 7-dimensional inverse sigma tensor of the distribution   
$P_{gce}(B,S,Q,E,\dots)$. Comparing this to a bivariate normal distribution   
Eq.(\ref{BND}), one finds:    
\begin{eqnarray} \label{term_A}  
\sum \limits_{j=1}^J ~ \lambda_{j,1}^2  &=& \frac{1}{\sigma_A^2 \left(1-\rho^2   
  \right)} ~=~ A  ~,\\ \label{term_B}  
\sum \limits_{j=1}^J ~ \lambda_{j,2}^2  &=& \frac{1}{\sigma_B^2 \left(1-\rho^2   
  \right)} ~=~ B  ~,\\ \label{term_C}  
\sum \limits_{j=1}^J ~ \lambda_{j,1} \lambda_{j,2} &=& -   
~\frac{\rho}{\left(1-\rho^2\right) \sigma_A \sigma_B}  ~=~ - C~.  
\end{eqnarray}   
After a short calculation one finds for the co-variances:  
\begin{eqnarray}   
\sigma_A^2 &=& \frac{B}{AB-C^2}~, \\   
\sigma_B^2 &=& \frac{A}{AB-C^2}~, \\   
\sigma_{A,B} &=& \frac{C}{AB-C^2}~,    
\end{eqnarray}   
and additionally the correlation coefficient, Eq.(\ref{rho}):   
\begin{eqnarray}   
\rho ~=~¸\frac{\sigma_{A,B}}{\sigma_A~\sigma_B} ~=~ \frac{C}{\sqrt{AB}}~,   
\end{eqnarray}   
where the terms $A,B,C$ are given by Eqs.(\ref{term_A} - \ref{term_C}).  
For the normalization in Eq.(\ref{PCE}) (from a comparison with   
Eq.(\ref{BND})) one finds:    
\begin{eqnarray}   
\frac{\det \sigma_N}{\det \sigma} &=& \frac{1}{\sigma_A \sigma_B   
  \sqrt{(1-\rho^2)}} ~=~ \sqrt{AB}~.  
\end{eqnarray}

\section{Acceptance Scaling}   
\label{AccS}  
To illustrate the `acceptance scaling` procedure employed in \cite{MCEvsData,CE_Res,SM_fluc_ce}  
we assume uncorrelated acceptance of particles of species $A$ and $B$.   
Particles are measured or observed with probability $q$ regardless of their momentum.  
The distribution of measured particles $n_A$, when a total number $N_A$ is produced,   
is then given by a binomial distribution:  
\begin{equation}  
P_{acc} \left(n_A|N_A \right)~=~ q^{n_A} ~\left(1-q \right)^{N_A-n_A}   
\binom{N_A}{n_A}~.  
\end{equation}  
The same acceptance distribution is used for particles of species $B$. Independent of the   
original multiplicity distribution $P(N_A,N_B)$, we define the moments of   
the measured particle multiplicity:  
\begin{equation}  
\langle n_A^a \cdot n_B^b \rangle ~\equiv~ \sum_{n_A,n_B} ~\sum_{N_A,N_B}~ n_A^a ~ n_B^b ~   
P_{acc} \left(n_A|N_A \right) ~ P_{acc} \left(n_B|N_B \right)~P(N_A,N_B)~.  
\end{equation}  
For the first moment $\langle n_A \rangle$ one finds:  
\begin{equation}  
\langle n_A \rangle ~=~ q~\langle N_A \rangle~.  
\end{equation}  
The second moment $\langle n_A^2 \rangle$ and the correlator   
$\langle n_A \cdot n_B \rangle$  are given by:  
\begin{eqnarray}  
\langle n_A^2 \rangle &=& q^2 \langle N_A^2 \rangle ~+~ q\left(1-q \right)  
\langle N_A \rangle~, \\  
\langle n_A \cdot n_B \rangle &=& q^2 \langle N_A \cdot N_B \rangle~.  
\end{eqnarray}  
For the scaled variance $\omega^A_q$ of observed particles one now finds \cite{SM_fluc_ce}:  
\begin{eqnarray}\label{acc_omega}  
\omega^A_q &=& \frac{\langle n_A^2 \rangle - \langle n_A \rangle^2}  
{\langle n_A \rangle} ~=~ 1~-~q~+q~\omega^A_{4\pi}~,  
\end{eqnarray}  
where $\omega^A_{4\pi}$ is the scaled variance of the distribution if all particles of   
species $A$ are observed. Lastly, the correlation coefficient $\rho_q$ is:  
\begin{equation}  
\rho_q ~=~ \frac{\langle \Delta n_A \Delta n_B \rangle}  
{\sqrt{\langle \left( \Delta n_A \right)^2 \rangle ~   
\langle \left( \Delta n_B \right)^2 \rangle  }}~,  
\end{equation}  
with $\langle \Delta n_A \Delta n_B \rangle = \langle n_A \cdot n_B \rangle -   
\langle n_A \rangle \langle n_B \rangle$, and $\langle \left( \Delta n_A \right)^2 \rangle  
= \langle n_A^2 \rangle - \langle n_A \rangle^2$.  
Substituting the above relations, one finds after a short calculation:  
\begin{equation}\label{acc_rho_comp}  
\rho_q ~=~ \rho_{4\pi}~q~\sqrt{\omega^A_{4\pi} \omega^B_{4\pi}}   
~\Big[~  
q^2\omega^A_{4\pi} \omega^B_{4\pi}   
~+~ q(1-q)\omega^A_{4\pi}  ~+~ q(1-q)\omega^B_{4\pi}   
~+~ (1-q)^2  
 ~\Big]^{-1/2}~.  
\end{equation}  
In case $\omega^A_{4\pi}= \omega^B_{4\pi}=\omega_{4\pi}$, Eq.(\ref{acc_rho_comp}) simplifies to:  
\begin{equation}\label{acc_rho}  
\rho_q ~=~ \rho_{4\pi}~\frac{q~\omega_{4\pi}}{1~-~q~+~q~\omega_{4\pi}}~.  
\end{equation}  
Both lines are independent of the mean values $\langle N_A \rangle$ and $\langle N_B \rangle$.


\end{document}